\documentclass[11pt,a4paper]{article}
\pdfoutput=1
\usepackage[pdftex]{graphics}
\usepackage{jheppub}
\usepackage{amsmath,amssymb,amsfonts}
\usepackage{enumitem}
\usepackage{multirow}
\usepackage{array,booktabs}
\usepackage{slashed}

\usepackage{accents}
\usepackage{mathrsfs}
\usepackage{mathtools}

\usepackage[export]{adjustbox}
\setlist[itemize]{
	label=\adjustbox{scale=0.7}{$\bullet$}, itemsep=-3pt,topsep=0px
}

\usepackage[usenames,dvipsnames]{xcolor}
\definecolor{labelcolor}{RGB}{194, 175, 116}
\definecolor{rmkcolor}{RGB}{15,120,255}

\usepackage[final]{showlabels} 

\newif\ifToggleMacros
\ToggleMacrostrue  
\ifToggleMacros
\newcommand{\jwk}[1]{\textbf{\color{blue}{[JW: #1]}}}
\newcommand{\jh}[1]{\textbf{\color[RGB]{224,26,187}{[Joonhwi: #1]}}}
\newcommand{\jwl}[1]{\textbf{\color[RGB]{100,185,30}{[Jungwon: #1]}}}

\else
\newcommand{\jwk}[1]{}
\newcommand{\jh}[1]{}
\newcommand{\jwl}[1]{}
\fi

\let\oldc\c
\let\oldi\i

\newcommand{\fref}[1]{Fig.\,\ref{#1}}
\renewcommand{\eqref}[1]{Eq.\,(\ref{#1})}
\newcommand{\eqrefs}[2]{Eqs.\,(\ref{#1}) and (\ref{#2})}

\newcommand{\rcite}[1]{ref.\,\cite{#1}}
\newcommand{\rrcite}[1]{Refs.\,\cite{#1}}
\DeclareMathOperator{\diag}{diag}

\def\Re{{\operatorname{Re}}}

\def\mem{\hspace{0.1em}}

\def\nem{\hspace{-0.1em}}
\def\hnem{\hspace{-0.05em}}

\def\a{\alpha}
\def\b{\beta}
\def\c{\gamma}

\def\ve{\varepsilon}
\def\m{\mu}
\def\n{\nu}
\def\r{\rho}
\def\s{\sigma}

\def\l{\lambda}
\def\t{\tau}
\def\bpsi{{\smash{\bar{\psi}}\kern0.02em\vphantom{\psi}}}

\def\mathe{e}

\let\oldcap\cap
\renewcommand{\cap}{{\,\oldcap\,}}

\def\Pexp{\mathrm{P}\kern-0.1em\exp}

\def\Pexp{\mathrm{P}\kern-0.1em\exp}

\def\F{{\mathcal{F}}}
\def\G{{\mathcal{G}}}

\def\P{{\mathcal{P}}}
\def\C{{\mathcal{C}}}

\def\U{{\mathcal{U}}}

\newcommand{\dbar}{
	d\kern-.20em\makebox[0pt][l]{$\bar{}$}\kern.20em
}
\newcommand{\deltabar}{
	\delta\kern-.20em\makebox[0pt][l]{$\bar{}$}\kern.20em
}

\def\R{\mathbb{R}}

\def\Uinv{U_\cen\kern-0.3em\mathrlap{\adjustbox{raise=0.11em}{$^{-1}$}}\kern0.95em}

\def\bpsi{\tilde{\psi}}

\let\oldparagraph\paragraph
\newcounter{alphnum}
\renewcommand{\paragraph}[1]{
	\refstepcounter{alphnum}{%
		\oldparagraph{\scshape\alph{alphnum}. #1}\phantom{}\vskip0.35ex\noindent
	}%
}
\counterwithin*{alphnum}{subsubsection}

\def\mdot{{\mem\cdot\mem}}
\def\mplus{{\mem+\mem}}
\def\mminus{{\mem-\mem}}

\def\sprime{{\mathrlap{\smash{{}^\prime}}{\hspace{0.05em}}}}

\def\volt{\phi}

\def\bz{\bar{z}}

\title{
	Twisted Feynman Integrals: from generating functions to spin-resummed post-Minkowskian dynamics
}

\author[a]{Joon-Hwi Kim}
\author[b,c]{Jung-Wook Kim}
\author[d]{Jungwon Lim}

\affiliation[a]{Walter Burke Institute for Theoretical Physics,\\ California Institute of Technology, Pasadena, CA 91125}

\affiliation[b]{Theoretical Physics Department, CERN,
	1211 Geneva 23, Switzerland}

\affiliation[c]{Max Planck Institute for Gravitational Physics (Albert Einstein Institute),\\
	Am M\"{u}hlenberg 1, D-14476 Potsdam, Germany}

\affiliation[d]{Max Planck Institute f\"{u}r Physik, Werner-Heisenberg-Institut,\\ Boltzmannstr. 8, 85748 Garching, Germany}

\emailAdd{joonhwi@caltech.edu}
\emailAdd{jung-wook.kim@cern.ch}
\emailAdd{wonlim@mpp.mpg.de}

\abstract{
    We propose to call a class of deformed Feynman integrals as \emph{twisted Feynman integrals}, where the integrand has an additional exponential factor linear in loop momenta. Such integrals appear in various contexts: tensor reduction of Feynman integrals, Fourier transform of Feynman integrals, and spin-resummed dynamics in post-Minkowskian gravity. First, we construct a mathematical framework that manifests the geometric interpretation of twisted Feynman integrals. Next, we generalise the standard mathematical tools for studying Feynman integrals for application to their twisted cousins, and explore their mathematical properties. In particular, it is found that (i) Symanzik polynomials are no longer homogeneous and become graded, (ii) twisted Feynman integrals fall under the class of exponential periods, and (iii) the geometry of the function space cannot be inferred from the leading singularity computed through the (generalised) Baikov parametrisation of twisted Feynman integrals.
}

\begin{document}
	
	\begin{flushright}
		\footnotesize
		CERN-TH-2025-264\\
        \footnotesize
        MPP-2025-235
	\end{flushright}
    \vskip5pt
	\maketitle
	\section{Introduction}

Quantum field theory is sometimes \emph{defined} by perturbation theory\textemdash as a sum over Feynman diagrams~\cite{Cvitanovic:1983eb,Veltman:1994wz,Weinberg:1996kw}. 
Feynman integrals collectively refer to the type of integrals encountered when evaluating Feynman diagrams, and methods for efficient evaluation of Feynman integrals form the core of high-precision computations in quantum field theory. The ``precision revolution'' of quantum field theory from the 2000's was spearheaded by the synthesis of various algebraic methods that simplify the evaluation of Feynman integrals~\cite{Passarino:1978jh,Chetyrkin:1981qh,Kotikov:1990kg,Bern:1994zx,Bern:1994cg,Laporta:2000dsw,Ossola:2006us,Henn:2013pwa}. 

The tensor integral generating function~\cite{Feng:2022hyg} is among the youngest of the algebraic tools introduced for the evaluation of Feynman integrals. The generating functions were introduced as a means of performing tensor reduction of Feynman integrals: removing loop momenta dependence in the numerator. The integrals can be viewed as deformed Feynman integrals, where the deformation is due to an exponential factor\footnote{We follow the conventions of ref.~\cite{Chen:2024bpf} for the deformation factor since the geometric interpretation is more transparent.} $e^{i \sum_I \alpha_I \cdot \ell_I}$ with the deformation parameters $\alpha_I^\mu$ paired to their corresponding loop momenta $\ell_I^\mu$. 
Another motivation for studying these types of integrals comes from the phenomenology of gravitational wave physics: studying spin-resummed post-Minkowskian dynamics~\cite{Kim:2024grz,Chen:2024bpf,Aoude:2025xxq}.

We propose to call such deformed Feynman integrals as \emph{twisted Feynman integrals},\footnote{Our usage of ``twisted'' is based on the geometric interpretation of the exponential deformation factor
(originated from ``twisted gauge transformation'' of \rcite{GenSymGrav}).
It should be clarified that
this usage differs from that of ``twisted'' in the twisted-cohomology-based approaches to Feynman integrals; in the latter, ``twisted'' refers to the deformation of the cohomology differential $d \mapsto d + \omega \, \wedge$.
} because this terminology better reflects the \emph{geometric structure} underlying the deformation. It is not uncommon in theoretical physics to find examples where the original algebraic idea is later found to have a geometrical understanding: supersymmetry and superspace~\cite{Wess:1974tw,Salam:1974yz}, modern numerical bootstrap and positive geometry~\cite{Rattazzi:2008pe,Arkani-Hamed:2018ign}, and integration-by-parts (IBP) relations and intersection theory~\cite{Chetyrkin:1981qh,Mastrolia:2018uzb}. 
Tensor integral generating functions and twisted Feynman integrals are the latest pair to join this long list of examples.

This manuscript is organised as follows. We give an overview of various motivations for considering twisted Feynman integrals and explain the underlying geometric interpretation in sec.~\ref{sec:MOT}.
Next, we provide a precise mathematical definition of twisted Feynman integrals
in sec.~\ref{sec:DEF},
in both rigorous 
and intuitive ways
via analogies with electric circuits in magnetic fields.
We then return to the original algebraic context in sec.~\ref{sec:REP} and study the algebraic properties that distinguish twisted Feynman integrals from the original Feynman integrals, e.g., the breakdown of homogeneity and introduction of grading in Symanzik polynomials, the values of integrals not being given by periods~\cite{Marcolli:2009zy} but rather by exponential periods, and the function space not being captured by leading singularities computed through the Baikov representation~\cite{Baikov:1996iu}. We give concluding remarks in sec.~\ref{sec:SUM}.

	\section{Physical Motivations}
\label{sec:MOT}

\emph{Twisted Feynman integrals} are deformations of typical Feynman integrals where the integrand is modified by an exponential factor. For example, an $L$-loop integral $I_L$ is deformed into the $L$-loop twisted integral $I_L [\a_1^\m, \cdots , \a_L^\m]$ by the substitution
\begin{align}
    I_L = \int_{\ell_1, \cdots , \ell_L} \hskip -15pt X (\ell_1, \cdots, \ell_L) \quad \mapsto \quad I_L [\a_1^\m, \cdots , \a_L^\m] = \int_{\ell_1, \cdots , \ell_L} \hskip -15pt X (\ell_1, \cdots, \ell_L) \,e^{i \sum_{n=0}^L \ell_n \cdot \a_n} \,, \label{eq:TFI_def}
\end{align}
where $X(\ell)$ is the original integrand. We review the motivations for considering such integrals (tensor reduction of Feynman integrals and spin-resummed dynamics in post-Minkowskian gravity) and provide a geometric interpretation associated to the exponential deformation factor.

Similar to Feynman integrals, twisted Feynman integrals should be viewed as generalised functions: they are defined by the consistency relations between the integrals implied by their integrands, unless the integral expression converges and has a well-defined value. We adopt this viewpoint to resolve the issue of well-definedness and finiteness of twisted Feynman integrals.

This introductory exposition will also set up basic notations.
We work in $D$ spacetime dimensions,
in Euclidean or Lorentzian signatures.
Spacetime indices are denoted with
$\m,\n,\r,\s,\cdots$. 
The $D$-dimensional $L$-loop momentum measure is denoted as $d^{DL}\ell$, which is sometimes abbreviated as subscripts of the integral sign; e.g., $\int_{\ell_1,\cdots,\ell_L}$.

\subsection{Generating Function for Tensor Integrals}
\label{sec:MOT.gf}

One motivation for considering twisted Feynman integrals is to define generating functions for tensor integrals~\cite{Feng:2022hyg}. For example, consider a sequence of one-loop integrals,
\begin{align}
		\label{eq:int.gens}
		I \,=
		\int d^D \ell\,\,\,
			\frac{1}{\mathcal{D}(\ell)}
		\,,\quad
		I_\m \,=
		\int d^D \ell\,\,\,
		\frac{
			\ell_\m
		}{\mathcal{D}(\ell)}
		\,,\quad
		I_{\m\n} \,=
		\int d^D \ell\,\,\,
		\frac{
			\ell_\m \ell_\n
			}{\mathcal{D}(\ell)}
		\,,\quad
		\cdots
		\,,
\end{align}
where $\mathcal{D}(\ell)$ is the denominator constructed as products of inverse propagators. $\mathcal{D}(\ell)$ is a polynomial of loop momentum $\ell \in \mathbb{R}^D$, and we have suppressed the dependence on external parameters for simplicity. This series of integrals can be packaged into a single expression
\begin{align}
	\label{eq:int.gen}
	I(\a)
	\,=
	\int d^D \ell\,\,\,
	\frac{e^{i \ell\cdot\a}}{\mathcal{D}(\ell)}
	\,,
\end{align}
which can be viewed as the generating function\footnote{In probability theory, such deformations are referred to as \emph{characteristic functions} rather than moment generating functions due to the extra imaginary unit $i$ in the exponent. The distinction is relevant in probability theory because of convergence, e.g., the Lorentzian distribution (Cauchy distribution) does not have a moment generating function but has a characteristic function. The distinction is irrelevant for twisted Feynman integrals since we understand them as generalised functions where the analytic continuation $\a \mapsto i \a$ is not obstructed.} for tensor integrals respect to the vector parameter $\a \in \R^D$: the coefficients of $I(\a)$ as a Taylor series in $\a$ are the tensor integrals, i.e.
\begin{align}
	I(\a)
	\,=\,
		I
		\,+\,
			\frac{i}{1!}\,
			\a^\m I_\m
		\,+\,
			\frac{i^2}{2!}\,
			\a^\m \a^\n\, I_{\m\n}
		\,+\,
			\cdots
	\,.
\end{align}
Note that this is a \emph{scalar} integral; $I(\a)$ does not have any free Lorentz indices.

If we can explicitly evaluate $I(\a)$ as a function of $\a$, we can compute tensor integrals of arbitrary rank by taking derivatives of $I(\a)$ with respect to $\a$. In the conventional approach to tensor integrals, the Feynman integrals are first projected onto the basis of integrals that belong to one of the IBP sectors and then reduced to master integrals through IBP relations. For high-rank tensor integrals\textemdash that can be encountered, e.g., in effective field theory (EFT) approaches to gravity\textemdash the projection to IBP sectors can become computationally demanding because of the large number of terms generated from the projection. One of the motivations for introducing tensor integral generating functions is to bypass this difficulty at the cost of extra vector parameters in the integral: as an alternative method for tensor reduction of Feynman integrals.

So far, we have described an \emph{algebraic} motivation for considering twisted Feynman integrals: as an algebraic tool for performing tensor reduction of Feynman integrals. 
In the next section, we give a brief overview of a physical setup in which twisted Feynman integrals naturally emerge, which alludes to their \emph{geometric} interpretation.

\subsection{Spinning Black Holes in Post-Minkowskian Gravity}
\label{sec:MOT.zigzag}

One ``peculiarity'' of the Kerr solution is that it can be constructed from the Schwarzschild solution by shifting
its center
in the imaginary direction $i a$, where $a = {S}/{M}$ is the spin-length four-pseudovector of the resulting black hole:
\begin{align}
    \label{pmia}
    x
    \quad\to\quad
    x \pm ia
    \,.
\end{align}
This solution-generating technique is
known as the \textit{Newman-Janis algorithm}.
The Newman-Janis algorithm is the very historical method that derived the Kerr-Newman black hole solution in 1965
\cite{Newman:1965my-kerrmetric}.
However, it has long been remained at the status of a mere ``trick,''
due to the lack of a clear physical nor geometrical origin.
There has been many attempts to explain the Newman-Janis algorithm:
geometrically \cite{talbot1969newman,Drake:1998gf,Gurses:1975vu,flaherty1976hermitian,grg207flaherty,Rajan:2016zmq,giampieri1990introducing,Newman:1965tw-janis,penrose1967twistoralgebra},
from a theory of complex spacetimes
\cite{newman1988remarkable,Newman:1973afx,Newman:2002mk,Newman:1973yu,newman1974curiosity,newman1974collection,newman1973complex,newman2004maxwell,newman1976heaven,ko1981theory,grg207flaherty,sst-asym},
and from scattering amplitudes
\cite{Arkani-Hamed:2017jhn,Guevara:2018wpp,Guevara:2019fsj,chkl2019,aho2020}.
See \cite{Adamo:2014baa,Erbin:2016lzq,nja} for historical reviews on this subject.

The original Newman-Janis algorithm concerns stationary solutions
in Einstein-Maxwell theory. 
A related question to ask is 
whether the imaginary shift of Newman and Janis
can be applied to \emph{dynamical} black holes: can the Newman-Janis shift be applied to the worldlines of black holes that experience kicks due to interactions with other sources? 
We remark that this question is more than an academic interest and has phenomenological applications.
Current waveform models used in LIGO-Virgo-KAGRA data analysis are known to suffer from systematic biases that grow as the binary constituents spin faster~\cite{Dhani:2024jja}. 
Although the biases were expected to become relevant for future-generation gravitational wave observatories, the systematic uncertainties of waveform models in the high-spin regime have already limited our capabilities to determine the source properties of gravitational waves: the constituent spin orientation could not be determined in the recent observation GW231123, which is expected to be the remnant from a merger of rapidly-spinning black holes (dimensionless spins $\chi_1 \sim 0.9$ and $\chi_2 \sim 0.8$ respectively)~\cite{LIGOScientific:2025rsn}. 
Since waveform models are built on our understanding of the two-body dynamics, we can expect to improve the waveform models in the high-spin regimes if we can study how spin effects resum in the two-body dynamics~\cite{Kim:2024grz,Chen:2024bpf}.
Indeed, there are indications that resumming spin effects lead to modifications of the singularities in the post-Minkowskian two-body dynamics beyond the leading order~\cite{Aoude:2022thd,Damgaard:2022jem,Kim:2024grz,Chen:2024bpf,Bohnenblust:2024hkw,Aoude:2025xxq}, and the modification provides justification of the effective-one-body (EOB) map $\vec{a}_{\text{eff}} = \vec{a}_1 + \vec{a}_2$ used in EOB-based waveform models such as \texttt{SEOBNRv5}~\cite{Ramos-Buades:2023ehm,Khalil:2023kep} and \texttt{SEOB-PM}~\cite{Buonanno:2024vkx} beyond the leading post-Newtonian order. However, it remains to be explored whether this EOB map can be justified beyond the leading order in the mass-ratio expansion.

The first hints that promoting the Newman-Janis shift into a dynamical statement is indeed possible were provided by on-shell approaches to scattering amplitudes: the Arkani-Hamed, Huang, and Huang (AHH) minimal coupling for massive higher-spin particles~\cite{Arkani-Hamed:2017jhn} was determined to reproduce the multipole moments of Kerr(-Newman) black holes~\cite{Guevara:2018wpp,Chung:2018kqs,Chung:2019yfs}, which was later understood to be an on-shell scattering amplitude realisation of the Newman-Janis shift~\cite{Arkani-Hamed:2019ymq}.
This understanding sparked two separate research directions for promoting the Newman-Janis shift into a dynamical statement, where the resulting spinning point particles should be understood as idealisations of Kerr black holes, since horizon absorption effects have been neglected.

The first is 
amplitudes-oriented,
adopting the minimal coupling criterion for three-point on-shell amplitudes as the ``definition'' of the Newman-Janis shift 
at the linearised order
and generalising the criterion to arbitrary multiplicities and off-shell currents, which is the strategy followed by massive higher-spin gauge symmetry~\cite{Chiodaroli:2021eug,Cangemi:2022bew} and heavy-mass/heavy-particle recursion relations~\cite{Bjerrum-Bohr:2023jau,Bjerrum-Bohr:2023iey}.
However, this approach has not quite clarified yet on the status of Newman-Janis shift at the nonlinear order,
due to the very notorious contact term ambiguity
in the Compton amplitude \cite{Arkani-Hamed:2017jhn}.

The other is 
Lagrangian-oriented,
applying the imaginary shift on the worldline of the spinning black hole 
in its effective point-particle description.
A proposal has been given by \rcite{Guevara:2020xjx},
which identified a worldsheet structure.
For instance,
\rcite{Kim:2023aff} has imported this proposal
into a twistorial framework 
to compute physical observables in the scattering problem of
the electromagnetic analogue \cite{Lynden-Bell:2002dvr,aho2020,Newman:1965tw-janis} of a Kerr black hole.
This twistor-theoretic approach has been initiated by
\rrcite{Kim:2021rda,Kim:2024grz},
building upon which
an explicit realisation of the worldsheet action
has been achieved \cite{probe-nj,sodual}
via the mathematical framework established in \rcite{gde}.
We further elaborate on this approach, since the approach advocates a viewpoint that provides a natural setup for the geometric interpretation of twisted Feynman integrals.

\begin{figure}
    \centering
    \includegraphics[scale=1]{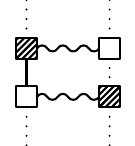}
    \caption{
        A one-loop diagram for spin-resummed dynamics that requires one-loop master integrals of the form \eqref{eq:TFI_2PM_ex}. Wavy lines denote massless mediator fields, dotted lines denote background worldline of massive particles, solid lines denote worldline fluctuations, and empty/shaded boxes denote MHV/$\overline{\text{MHV}}$-type couplings with the massless mediator field. 
        Figure reproduced from Ref.~\cite{Kim:2024grz}.
    }
    \label{fig:zigzags}
\end{figure}

\fref{fig:zigzags} represents a typical diagram in the twistor-theoretic approach to the dynamical Newman-Janis shift. 
The empty/shaded boxes denote couplings of the black hole to massless force carriers
in the positive/negative helicity (self-dual/anti-self-dual) sector.
Crucially,
the positive-helicity coupling is realised at a complex point $z = x - ia$,
whereas the negative-helicity coupling is realised at a complex point $\bz = x + ia$.
These localizations at the positive- and negative-helicity points
are shown to exactly describe the Newman-Janis algorithm
at the level of the linear-in-force-carrier (three-point) coupling
\cite{aho2020},
and an on-shell diagram approach
has been considered by \rcite{sst-asym}
in terms of complexified massive spinning kinematics.

The complex points $z$ and $\bz$ can be taken more literally; a Kerr black hole can be represented as a pair of self-dual and anti-self-dual gravitational instantons~\cite{hawking1977gravitational,Gibbons:1978tef} (extremal cases of complexified Taub-NUT solutions \cite{Taub:1950ez,Newman:1963yy}), each located at complexified spacetime points $\bz$ and $z$. This is not a statement about linearised gravity; the full \emph{nonlinear} Kerr solution can be represented as a pair of self-dual and anti-self-dual gravitational instantons~\cite{nja}, which can be proved based on the ideas of magnetic monopoles~\cite{Taub:1950ez,Newman:1963yy,Misner:1963flatter,Bonnor:1969ala,sackfield1971physical,demianski1966combined,Plebanski:1975xfb,dowker1974nut,Griffiths:2009dfa}
and classical double copy~\cite{monteiro2014black}. The instantons are separated by the displacement $z - \bz = - 2 ia$, therefore the Kerr black hole is viewed as an extended (nonlocal) object constructed from a pair of point particles. This nonlocality generates the twist in Feynman integrals encountered in spin-resummed post-Minkowskian dynamics. We consider one-loop integrals as an example.

Typical one-loop integrals in spin-resummed post-Minkowskian dynamics can be evaluated from the following class of one-loop master integrals ($v \cdot q = 0$)
\begin{align}
    I_{\l_1,\l_2,\l_3} &= \int \frac{d^D \ell \, e^{2 \ell \cdot a}}{(\ell^2)^{\l_1} [(q - \ell)^2]^{\l_2} (2 v \cdot \ell - i0^+)^{\l_3}} \,, \label{eq:TFI_2PM_ex}
\end{align}
which has been evaluated for generic values of propagator exponents ($\vec{\l}$) in \rcite{Kim:2024grz}. The appearance of the exponential factor $e^{2 \ell \cdot a}$ in the integrand\textemdash which makes the integral a twisted Feynman integral [\eqref{eq:TFI_def}]\textemdash 
is the telltale sign of the Newman-Janis shift: a finite displacement by $x$ corresponds to the factor $e^{i p \cdot x}$ in momentum space, and the factor $e^{2 \ell \cdot a} = e^{i \ell \cdot (- 2 i a)}$ corresponds to a finite separation of $-2 i a$, which is exactly the displacement between the gravitational instantons representing a Kerr black hole.

In the next subsection,
we elaborate on the interpretation of the twist as a nonlocality,
by considering the twisting of a scalar particle amplitude
as a concrete example.

\subsection{Twisted Feynman integrals in Position Space}
\label{sec:MOT.jump}

Although Feynman integrals are usually written in momentum space, the geometric interpretation of twisted Feynman integrals becomes clear when the integral is written in position space, where the integrand is given by products of Green's functions and we integrate over the positions of internal vertices. A Feynman integral given in momentum space is converted to its corresponding position space representation through the following steps, where we employ shorthand notations
$\dbar^D\ell := {d^D\ell}/{(2\pi)^D}$
and
$\deltabar^{(D)}(\ell) := (2\pi)^D\, \delta^{(D)}(\ell)$ to avoid factors of $2\pi$ in the equations:
\begin{enumerate}
    \item Deconstruct momentum conservation conditions imposed on each vertex: label all internal edges by its respective momentum $\ell_j$ and give each vertex the momentum conservation condition $\deltabar^{(D)} (\sum_{\text{in}} \ell_j^\m - \sum_{\text{out}} \ell_j^\m)$ (include external momentum if external legs are also attached to the vertex).
    \item Label each vertex by its respective position $x_k$ and (inverse) Fourier transform the momentum conservation condition to $e^{i \sum_{\text{in}} x_k \cdot \ell_j - i \sum_{\text{out}} x_k \cdot \ell_j}$.
    \item Perform all (inverse) Fourier transform associated to the momenta of internal edges. If the momentum $\ell_i$ associated to the edge $i$ flows from the vertex $x_j$ to the vertex $x_k$, the edge $i$ becomes the Green's function in position space $G (x_k - x_j \,;\, 0) = G (x_k \,;\, x_j)$, having the interpretation of a particle propagating from the point $x_j$ to $x_k$.
    \item Remove the volume factor associated to introducing the external momentum conservation conditions. This is equivalent to choosing an internal vertex and fixing its coordinates as the origin, while dropping the overall momentum conservation condition $\deltabar^{(D)} (\sum_{\text{in}}p_j^\m - \sum_{\text{out}} p_j^\m)$ of external momenta.
\end{enumerate}
The end result is an integral where external momenta are associated to the plane wave factors $e^{i p\cdot x}$, all internal vertices (except for the vertex fixed as the origin) are mapped to integrations in position space $\int d^D x_j$, and all edges are mapped to propagators $G (x_j \,;\, x_k)$ between the vertices. The set of consecutive edges that form a loop passing through the vertices $\{ x_{j_0} ,\, x_{j_1} ,\, x_{j_2} ,\, \cdots ,\, x_{j_0} \}$ has an interpretation as the worldline of a particle that makes a round trip consisting of spacetime points $\{ x_{j_0} ,\, x_{j_1} ,\, x_{j_2} ,\, \cdots ,\, x_{j_0} \}$, which is the interpretation behind the string-inspired formalism (also known as the worldline formalism) for amplitudes~\cite{Schubert:2001he}.

Select an edge of momentum $\ell$, which flows from the vertex $x_j$ to the vertex $x_k$. This edge corresponds to the Green's function $G (x_k \,;\, x_j )$. The deformation of the Feynman integral by a factor $e^{i \alpha \cdot \ell}$ is equivalent to substituting this Green's function by $G (x_k + \alpha \,;\, x_j)$, and the loop containing this edge now has an interpretation as the worldline of a particle that makes a ``twisted'' round trip; the particle ends up being shifted by $\alpha$ from the original starting point after the round trip.

\begin{figure}
	\centering
	\includegraphics{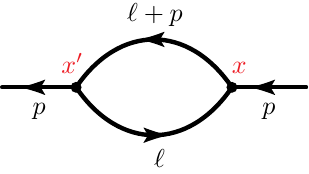}
	\qquad\qquad
	\includegraphics{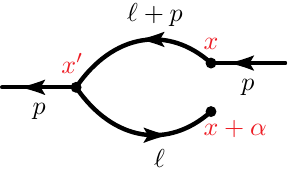}
	\caption{
		A one-loop Feynman graph and its twist,
		with momentum and position labels.
	}
	\label{fig:peng}
\end{figure}

As a concrete example, let us consider the one-loop massless bubble integral depicted in the first diagram of \fref{fig:peng}. The diagram represents the Feynman integral
\begin{align}
	\label{eq:peng}
	I(p) \,=
		\int \dbar^D\ell\,\,\,
			\frac{1}{\ell^2}
			\frac{1}{(\ell+p)^2}
	\,,
\end{align}
where $p \in \mathbb{R}^D$ is the external momentum. Following the outlined steps, the integral is represented in position space as
\begin{align}
\begin{aligned}
    I (p) \,&= \int d^D x \, e^{i p \cdot x} \left( \int \dbar^D \ell \, \frac{e^{i \ell \cdot x}}{\ell^2}\right) \left( \int \dbar^D \ell' \, \frac{e^{i \ell' \cdot (-x)}}{\ell'^2}\right)
    \\ &= \int d^D x \, e^{i p \cdot x} \, G(x,0) \, G(-x,0) \, = \int d^D x \, e^{i p \cdot x} \, G(x,0) \, G(0 , x) \,,
\end{aligned} \label{eq:peng2}
\end{align}
where we have fixed $x' = 0$ as the origin to remove the total momentum conservation factor
and $G(x,y)$ is the Green's function in position space,
\begin{align}
    G (x,y) &= \int \dbar^D \ell \, \frac{e^{i \ell \cdot (x-y)}}{\ell^2} \,,
\end{align}
which satisfies $G(x,y) = G(x-y,0)$ due to translation invariance. The interpretation of the last expression in \eqref{eq:peng2} is clear; the virtual particle in the loop makes a round trip from $x$ to $x' = 0$ (represented by $G(0,x)$), and then back to $x$ (represented by $G(x,0)$).

We now consider the ``twisting'' of \eqref{eq:peng} by the factor $e^{i \ell \cdot \a}$, which is represented by the second diagram of \fref{fig:peng}:
\begin{align}
\begin{aligned}
    \label{eq:tpeng}
	I(p;\a) \,&=
		\int \dbar^D\ell\,\,\,
			e^{i\ell\cdot\a}\,
			\frac{1}{\ell^2}
			\frac{1}{(\ell+p)^2}
	\\ &= \int d^D x \, e^{i p \cdot x} \left( \int \dbar^D \ell \, \frac{e^{i \ell \cdot (x + \a)}}{\ell^2}\right) \left( \int \dbar^D \ell' \, \frac{e^{i \ell' \cdot (-x)}}{\ell'^2}\right)
    \\ &= \int d^D x \, e^{i p \cdot x} \, G(x + \a,0) \, G(-x,0) \, = \int d^D x \, e^{i p \cdot x} \, G(x + \a,0) \, G(0 , x) \,.
\end{aligned}
\end{align}
Note that the first Green's function has been changed from $G (x,0)$ to $G (x+\a,0)$. This time, the virtual particle makes a round trip from $x$ to $x' = 0$ ($G(0,x)$), but returns to $x + \a$ instead of $x$ ($G(x+\a,0)$); the loop made by the virtual particle is now ``twisted open.'' This is the geometric interpretation of twisted Feynman integrals: the deformation of the integrand by the factor $e^{i \ell \cdot \a}$ ``twist opens'' the loop made by the virtual particle corresponding to the loop momentum $\ell$, and the virtual particle becomes displaced by $\a$ after the round trip. 

A final remark is that the interpretation of ``where'' the loop has been twisted open is not unique. For example, we could have rewritten the last line of \eqref{eq:tpeng} as
\begin{align}
    I (p;\a) &= \int d^D x \, e^{i p \cdot x} \, G(\a, -x) \, G(-x,0) \,,
\end{align}
which now has an interpretation as a round trip from $x' = 0$ to $-x$ and then returning to $\a$, opening up the loop at $x'=0$ rather than at $x$. 
In the next section, we construct a mathematical framework that encompasses all possible geometric interpretations of twisted Feynman integrals.

\section{Mathematical Definitions}
\label{sec:DEF}

Taming infinities from Feynman integrals is a well-known difficulty of quantum field theory, and one adopts various regularisation schemes to systematically remove divergences appearing in the integrals. The most widely used regularisation scheme is the \emph{dimensional regularisation} scheme, where the spacetime dimension $D$ is considered as a complex variable and computations are organised as Laurent expansions of $D$ around the physical spacetime dimension, e.g. $D = 4 - 2 \varepsilon$ with $\ve$ as the Laurent expansion parameter.

One of the axioms of dimensional regularisation is \emph{translation invariance} of loop momentum, where shifting the loop momentum $\ell$ appearing in the loop integrand $f(\ell)$, e.g. $f(\ell) \mapsto f (\ell+p)$, does not change the value of the integral;\footnote{We will use ``translation invariance'' to exclusively refer to this property throughout the manuscript.} see e.g. \textsection 4.1 of Ref.~\cite{Collins:1984xc}. On the other hand, the deformation to twisted Feynman integrals clearly does not respect this axiom. Consider the one-loop bubble integral given as an example in sec.~\ref{sec:MOT.jump}; while translation invariance implies the equivalence of the integrals
\begin{align}
    \int d^D\ell\,\,\,
			\frac{1}{\ell^2}
			\frac{1}{(\ell+p)^2} 
    \,\,\,
    \,=\,
    \,\,\,
            \int d^D\ell\,\,\,
			\frac{1}{(\ell-p)^2}
			\frac{1}{\ell^2} \,,
\end{align}
the deformations of the integrals to twisted Feynman integrals are \emph{inequivalent}, i.e.
\begin{align}
    \int d^D\ell\,\,\,
			e^{i\ell\cdot\a}\,
			\frac{1}{\ell^2}
			\frac{1}{(\ell+p)^2} 
        \,\,\,
        \,\neq\,
        \,\,\,
        \int
            d^D\ell\,\,\,
			e^{i\ell\cdot\a}\,
			\frac{1}{(\ell-p)^2}
			\frac{1}{\ell^2} \,. \label{eq:TIbreakingTFI}
\end{align}
However, they are not entirely unrelated since they only differ by the overall factor $e^{i p \cdot \a}$, and \emph{for the purpose of studying the space of functions} it would be more advantageous to consider the two as equivalent.

This observation raises a natural question: is it possible to define twisted Feynman integrals in a way that maintains the translation invariance of Feynman integrals? More concretely, since twisted Feynman integrals have a geometric interpretation as opening up the loops formed from the worldline trajectories of virtual particles, is it possible to construct a geometric framework for (twisted) Feynman integrals where the equivalence relation $\sim$ for integral expressions derived from the framework implies equivalence of the LHS and RHS of \eqref{eq:TIbreakingTFI}, i.e.
\begin{align}
    \int d^D\ell\,\,\,
			e^{i\ell\cdot\a}\,
			\frac{1}{\ell^2}
			\frac{1}{(\ell+p)^2} 
    \,\,\,
    \,\stackrel{?}{\sim}\,
    \,\,\,
            \int d^D\ell\,\,\,
			e^{i\ell\cdot\a}\,
			\frac{1}{(\ell-p)^2}
			\frac{1}{\ell^2} \,.
\end{align}
We answer this question in the affirmative through the geometric language of (co)homology on graphs \cite{Hatcher}. The key observation is that translation invariance of Feynman integrals associate a loop momentum $\ell^I$ to the corresponding \emph{cycle} $\mathcal{C}_I$ in the Feynman diagram, rather than to an internal \emph{edge} (or a propagator) of it.

\subsection{Twisted Feynman Graph as a Feynman Graph in Magnetic Field}

Our construction is based on the observation that loop momenta $\ell^I$ are associated to their corresponding loops (cycles) $\mathcal{C}_I$ of the graph. We will set up notation for graphs and provide an example to familiarise oneself to the idea: electric circuits. 

We consider a directed graph $\G$, which can be defined by the set of vertices $V = \{ v_i \}$ and the set of edges $E = \{ e_i \}$. Each edge $e_i$ corresponds to an ordered pair of vertices $e_i \leftrightarrow (v_{i_1}, v_{i_2})$ and carries an associated label, where the order of vertices denotes the direction of the edge (from $v_{i_1}$ to $v_{i_2}$). Each vertex also carries an associated label. We use $L$ to denote the number of loops, which means that the graph has $L$ independent cycles $\C_I$ ($I {\:=\:} {1,\cdots L})$. The cycles $\C_I$ form a basis for the first homology group $H_1(\G)$ over integers.

When the graph $\G$ is considered as a Feynman graph, the label associated to the edge $e$ is the $D$-dimensional momentum $p_{e} \in \mathbb{R}^D$ and the label associated to the vertex $v$ is the $D$-dimensional total external momentum flowing into the vertex $p_{v} \in \mathbb{R}^D$; for the electric circuit example considered in this section, the label associated to the edge $e$ is the current $i_e \in \mathbb{R}$ and the label associated to the vertex $v$ is the total external current flowing into the vertex $i_v \in \mathbb{R}$. 

Having set up the notation, we now proceed to the example of resistance-less electric circuits.
The reader may jump to the next section if the reader is willing to accept that a loop momentum should be associated to its corresponding cycle, rather than to an edge of the graph.

An electric circuit is a directed graph $\mathcal{G}$ with labeled edges, where the label $i_e \in \mathbb{R}$ denotes the current of the edge $e$. 
Current conservation demands that at each vertex $v_j$ the sum of currents flowing into the vertex must add up to zero,
\begin{align}
    i_{v_j} + \sum_{\substack{e \in E \\ e = (\bullet, v_j)}} i_e - \sum_{\substack{e \in E \\ e = (v_j,\bullet)}} i_e = 0 \,,
\end{align}
where $i_{v_j}$ is the external current flowing into the vertex $v_j$ (which is negative if the external current flows out of the vertex). Summing over all conservation conditions at each vertex demands that the external currents add up to zero $\sum_{v \in V} i_v = 0$.

\begin{figure}
    \centering
    \includegraphics[scale=0.85]{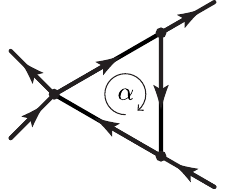}
    \caption{
        A twisted Feynman graph is 
        a Feynman graph
        put in
        a ``magnetic field.''
    }
    \label{fig:trifeyn}
\end{figure}

Now suppose an electric circuit is placed in an external time-varying magnetic field $\mathbf{B}$.
Each loop $\C_I$ in the circuit experiences
the induced electromotive force
$\oint_{\C_I} \mathbf{E} \mdot d\mathbf{x}$,
which equals the flux of $-\dot{\mathbf{B}}$
on account of Faraday's law.
Here, the index $I$ runs through $1$ to $L$,
where $L$ denotes the number of loops in the circuit.

For the sake of concreteness, we have illustrated a simple electric circuit that has three edges
in \fref{fig:trifeyn}.\footnote{The four external legs are not edges in our definition of a graph; they are labels corresponding to external currents that flow into the vertices they are attached to.}
This circuit has only one loop, say $\C_1$.
It is placed in a time-varying magnetic field
such that the loop is subject to the net electromotive force 
$\oint_{\C_I} \mathbf{E} \mdot d\mathbf{x} = \a$.

Now let us think about
the voltage drop 
\smash{$\volt_e = \int_e \mathbf{E} \mdot d\mathbf{x}$}
for each segment $e$ in the loop $\C_1$.
Certainly, the voltage drops
for the three edges of the graph in \fref{fig:trifeyn}
must add up to the net electromotive force, $\a$.
However, this information itself does not completely determine the voltage drops for each segment:
see a textbook discussion \cite{griffiths2023introduction}
and also \rcite{romer1982voltmeters}.\footnote{
    The indeterminacy arises due to the assumption of zero resistance;
    the voltage drops are fully determined
    when each segment of the circuit has a finite resistance.
}

\begin{figure}
    \centering
    $\begin{array}{ccc}
        \includegraphics[scale=0.85]{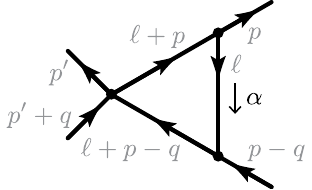}
        \quad&\quad
        \includegraphics[scale=0.85]{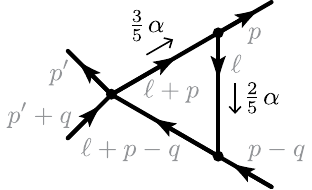}
        \quad&\quad
        \includegraphics[scale=0.85]{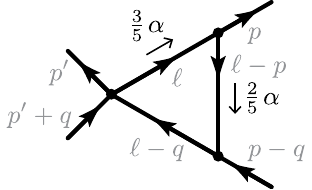}
        \\
        \text{\small (a)}
        &
        \text{\small (b)}
        &
        \text{\small (c)}
    \end{array}$
    \caption{Three realisations of the same twisted Feynman graph.}
    \label{fig:trifeyn-r}
\end{figure}

With this understanding,
\fref{fig:trifeyn-r}
provides 
three possible \textit{realisations} 
of the circuit in \fref{fig:trifeyn}.
By realisation, it means to completely specify both the current labels and voltage drops for all edges (labels on edges),
given a fixed choice of the external currents (labels on vertices).
The realisations (a) and (b) illustrate that
the voltage drops can be redistributed freely
as long as the net electromotive force for the loop is kept the same.
The realisations (b) and (c) illustrate that
there exists an ambiguity in assigning the current variables 
for each segment in a loop,
although the loop current $\ell$ remains the same.

An observation can be made at this point.
Consider the total power generated
from the internal edges:
\begin{align}
    P
    \,=
    \sum_{e\:\text{internal}}
        \volt_e \cdot i_e
    \,.
\end{align}
For the first realisation in \fref{fig:trifeyn-r},
the total power evaluates to
\begin{subequations}
\begin{align}
    \label{power-a}
    P_\text{(a)}
    \,=\,
    0 \cdot (\ell+p)
    + \alpha \cdot \ell 
    + 0 \cdot (\ell+p-q)
    \,=\,
        \alpha \cdot \ell
    \,.
\end{align}
On the other hand, for the second realisation in \fref{fig:trifeyn-r}, the total power evaluates to 
\begin{align}
    \label{power-b}
    P_\text{(b)}
    \,=\,
    \frac{3}{5}\mem \alpha \cdot (\ell+p)
    + \frac{2}{5}\mem \alpha \cdot \ell
    + 0 \cdot (\ell+p-q)
    \,=\,
        \alpha \cdot \ell
        + \frac{3}{5}\, \alpha \cdot p
    \,.
\end{align}
Meanwhile, for the third realisation in \fref{fig:trifeyn-r}, the total power evaluates to 
\begin{align}
    \label{power-c}
    P_\text{(c)}
    \,=\,
    \frac{3}{5}\mem \alpha \cdot \ell
    + \frac{2}{5}\mem \alpha \cdot (\ell-p)
    + 0 \cdot (\ell-q)
    \,=\,
        \alpha \cdot \ell - \frac{2}{5}\, \alpha \cdot p
    \,.
\end{align}
\end{subequations}
Eqs.\,(\ref{power-a})-(\ref{power-c})
clearly demonstrate that the definition of the power is ambiguous,
due to the very ambiguity in the realisations of the circuit in the time-varying magnetic field.
However, observe that the term $\ell \mdot \alpha$ is present in all cases,
so all computations agree 
up to terms that have no dependency in the loop current $\ell$.
We may refer to such terms as external factors.

In a precise language,
this means that 
the power computed in 
Eqs.\,(\ref{power-a})-(\ref{power-c})
all belong to the same equivalence class,
if we declare an equivalence relation as follows:
\begin{align}
    \label{ext-equiv}
    P \,\sim\, P'
    \quad\iff\quad
    \text{$P - P'$\: is an external factor}
    \,.
\end{align}

\subsection{Definition of Twisted Feynman Graph}

A Feynman graph is a directed graph $\mathcal{G}$ with labeled edges, where the label $p_e \in \mathbb{R}^D$ denotes the momentum of the edge $e$. 
Momentum conservation demands that at each vertex $v_j$ the sum of momenta flowing into the vertex must add up to zero,
\begin{align}
    p_{v_j} + \sum_{\substack{e \in E \\ e = (\bullet, v_j)}} p_e - \sum_{\substack{e \in E \\ e = (v_j,\bullet)}} p_e = 0 \,,
\end{align}
where $p_{v_j}$ is the external momentum flowing into the vertex $v_j$ (which is negative if the external momentum is flowing out of the vertex). Summing over all conservation conditions of each vertex demands that the external momenta add up to zero $\sum_{v \in V} p_v = 0$.

A cycle $\C_I$ of a Feynman graph is associated with the loop momentum $\ell^I{}_\m$.\footnote{The Lorentz indices were restored for clarity in this paragraph.}
The index notation here implies that
the loop momenta transform as 
$\ell^I{}_\m \mapsto (U^{-1})^I{}_J\mem \ell^J{}_\m$
if one changes the basis for the cycles as
$\C_I \mapsto \C_J\mem U^J{}_I$
with $U \in \mathrm{GL}(L,\mathbb{Z})$.
A twisted Feynman graph $(\G,\a)$ is a Feynman graph $\G$
endowed with an object of the index form $\a^\m{}_I$,
which we call the twist.
Again, the index structure 
fully specifies
what type of object this is.
Practically,
it is
simply
a $D {\,\times\,} L$ matrix.

A realisation of a Feynman graph $\G$
is a directed graph $\G$
where each edge $e$ is labeled with
a momentum $p_e \in \mathbb{R}^D$
such that
the momentum looping around the cycle $\C_I$ is $\ell^I$.
Realisations of a Feynman graph 
exhibit the ambiguity of momentum relabelings
that leave the loop momenta $\ell^I$ invariant.

A realisation of a twisted Feynman graph $(\G,\a)$
is a directed graph $\G$
where each edge $e$ is labeled with
a momentum $p_e \in \mathbb{R}^D$
and a vector $\volt_e \in \mathbb{R}^D$
such that
\smash{$\sum_{e\in \C_I} \volt_e = \a_I$}
and the momentum looping around the cycle $\C_I$ is $\ell_I$.
Realisations of a twisted Feynman graph 
exhibit not only the ambiguity of momentum relabelings
but also the ambiguity of redistributing the vectors $\volt_e$
as voltage drops,
such that each twist $\a_I$,
as the net electromotive force for each cycle $\C_I$,
is left invariant.

Finally,
we define the canonical pairing 
$\langle \a , \ell \rangle$
between the loop momenta $\ell$
and the twist $\a$ as the sum
\begin{align}
    \label{eq:can-pairing}
    \langle \a , \ell \rangle
    \,=\,
        \alpha^\m{}_I\mem \ell^I{}_\m
    \,.
\end{align}
We have used the index notation on the right-hand side, 
so the sums over $\m=1,2,\cdots,D$ and $I=1,2,\cdots,L$ are implied.
\eqref{eq:can-pairing} performs a complete index contraction between an $\mathbb{R}^D$-valued homology element and an $\mathbb{R}^D$-valued cohomology element,
so it is invariant under basis transformations in $\mathbb{R}^D$ as well as basis transformations in the space of cycles (the homology group $H_1(\G)$).

A realisation of the canonical pairing $\langle \a , \ell \rangle$ is the sum $\sum_e \volt_e \mdot p_e$ for all internal edges.
This is the total power generated by $p_e$ and $\volt_e$ as currents and voltage drops.
It is not difficult to see that any realisation satisfies
\begin{align}
    \label{eq:vp=la}
    \sum_e \volt_e \mdot p_e 
    \,=\,
    \langle \a , \ell \rangle + (\text{external factors})
    \,,
\end{align}
which is simply the general statement of the observation that we have made in 
Eqs.\,(\ref{power-a})-(\ref{power-c}).
Provided the equivalence relation in \eqref{ext-equiv},
\eqref{eq:vp=la} can be stated as
\begin{align}
    \bigg[\,{
        \sum_e \volt_e \mdot p_e 
    }\,\bigg]
    \,=\,
        [\mem{ \langle \a , \ell \rangle }\mem]
    \,,
\end{align}
where the square bracket denotes the equivalence class.

\subsection{Definition of Twisted Feynman Integral}

A realisation of a Feynman graph $\G$ uniquely defines a Feynman integral $I(\G)$ via Feynman rules, e.g. for scalars, 
\begin{align}
    \G \quad \to \quad I(\G) = \int_{\ell} \, \frac{1}{\prod_e (\mathcal{D}_e (\ell))^{\n_e}} \,,
\end{align}
where $e {\:=\:} {1,2,\cdots m}$ labels the edges,
$\mathcal{D}_e(\ell)$ are the propagators corresponding to the respective edges,
and $\int_\ell$ is the loop momenta integral.

Similarly, a realisation of a twisted Feynman graph $(\G,\a)$
uniquely defines a realisation $I_\a(\G)$ of a twisted Feynman integral.
This multiplies the integrand of the Feynman integral $I(\G)$ by the realisation of the exponential factor $\exp(i \langle \a , \ell \rangle )$, e.g. for scalars, 
\begin{align}
    (\G , \a) \quad \to \quad I_\a (\G) = \int_\ell\,\,
            \frac{
                \mathe^{i\langle\a,\ell\rangle}
            }{
                \prod_e
                    (\mathcal{D}_e(\ell))^{\n_e}
            } \,.
\end{align}

The realisations of a twisted Feynman integral are not unique, which is precisely inherited down from the non-uniqueness of the twisted Feynman graph $(\G,\a)$.
As a result, the non-uniqueness in the realisations of a twisted Feynman integral is given by the equivalence relation
\begin{align}
    \label{eq:Isim}
    I_\a(\G) 
    \,\,\,\sim\,\,\,
    I_\a(\G)\mem f_\text{ext}(\a,\G)
    \,,
\end{align}
where $f_\text{ext}(\a,\G)$ depends solely on the external momenta and the realisations of the twist.
In other words, choosing a different realisation induces a multiplicative external factor \textit{outside} the integral, which can be determined without evaluating the loop integrals. 

From the equivalence relation of twisted Feynman integrals [\eqref{eq:Isim}], we define the equivalence class of twisted Feynman integrals $[I_\a(\G)]$ as the set of all possible realisations of $I_\a(\G)$. 
Consequently, a twisted Feynman graph $(\G,\alpha)$ has a one-to-one correspondence to the equivalence class of twisted Feynman integrals $[I_\a(\G)]$. 

The study of function space for twisted Feynman integrals can be formulated as the study of function space for the equivalence class of twisted Feynman integrals, 
since each realisation in the equivalence class only differs by an overall multiplicative factor that can be determined without evaluating the loop integrals.
In other words, when given the problem of evaluating (a realisation of) a twisted Feynman integral, 
we may choose a realisation within its equivalence class that is the most convenient for the explicit evaluation of the integral, 
and the original integral can be determined by determining the multiplicative factor that relates the two realisations. 
The abstract formulation of twisted Feynman integrals as equivalence classes of integrals (realisations) provides a setup where the described approach to their evaluation becomes natural.

\subsection{A three-loop Example}

\begin{figure}
    \centering
    $\begin{array}{ccc}
        \includegraphics[scale=0.55]{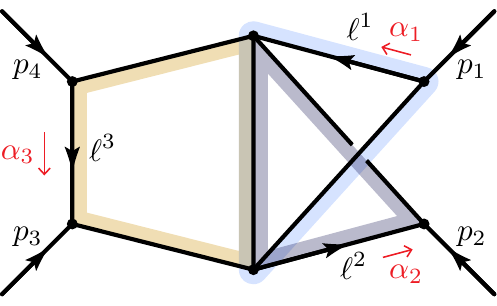}
        \,\,\,&\,\,\,
        \includegraphics[scale=0.55]{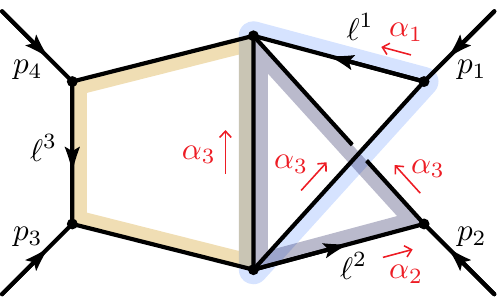}
        \,\,\,&\,\,\,
        \includegraphics[scale=0.55]{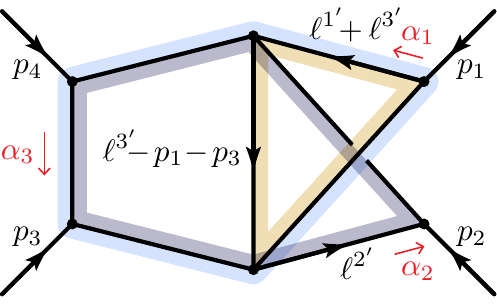}
        \\
            \text{\small (a)}
            &
            \text{\small (b)}
            &
            \text{\small (c)}
    \end{array}$
    \caption{
        Three realisations of the same twisted Feynman graph.
    }
    \label{fig:DX6}
\end{figure}

Let us end with a concrete demonstration to ensure that the ideas introduced in this section are conveyed in an explicit fashion.
\fref{fig:DX6} displays three different realisations of a single twisted Feynman graph with four external edges and three cycles (loops).
The realisations (a) and (b) illustrate the ambiguity in realising the twist: the redistribution of voltage drops.
The realisations (a) and (c) illustrate the freedom to choose different bases for the space of cycles.
Specifically, the cycles $\C_1, \C_2, \C_3$ in the realisation (a) are related to the cycles $\C_{1'}, \C_{2'}, \C_{3'}$ in the realisation (c) as
\label{matr}
\begin{align}
    \label{matr-C}
    \Big(\,{
        \C_{1'}
        \,\,\,
        \C_{2'}
        \,\,\,
        \C_{3'}
    }\,\Big)
    \,\,=\,\,
    \Big(\,{
        \C_{1}
        \,\,\,
        \C_{2}
        \,\,\,
        \C_{3}
    }\,\Big)
    \nem
    \left(\,\,
    \begin{matrix}
        \,\mathclap{1}\,
        \:\:&\:\:
        \,\mathclap{0}\,
        \:\:&\:\:
        \,\mathclap{1}\,
        \\
        \,\mathclap{0}\,
        \:\:&\:\:
        \,\mathclap{1}\,
        \:\:&\:\:
        \,\mathclap{0}\,
        \\
        \,\mathclap{1}\,
        \:\:&\:\:
        \,\mathclap{1}\,
        \:\:&\:\:
        \,\mathclap{0}\,
    \end{matrix}
    \,\,\right)
    \quad\leftrightarrow\quad
    \C_{I'} \,=\, \C_I\mem U^I{}_{I'}
    \,.
\end{align}
This implies that the loop momenta in the two realisations are related as
\begin{align}
    \label{matr-l}
    \begin{pmatrix}
        \ell^{1'}
        \\
        \ell^{2'}
        \\
        \ell^{3'}
    \end{pmatrix}
    \,\,=\,\,
    \left(\,\,
    \begin{matrix}
        \,\mathclap{0}\,
        \:\:&\:\:
        \,\mathclap{-1}\,
        \:\:&\:\:
        \,\mathclap{1}\,
        \\
        \,\mathclap{0}\,
        \:\:&\:\:
        \,\mathclap{1}\,
        \:\:&\:\:
        \,\mathclap{0}\,
        \\
        \,\mathclap{1}\,
        \:\:&\:\:
        \,\mathclap{1}\,
        \:\:&\:\:
        \,\mathclap{-1}\,
    \end{matrix}
    \,\,\,\,\right)
    \nem\hnem
    \begin{pmatrix}
        \ell^{1}
        \\
        \ell^{2}
        \\
        \ell^{3}
    \end{pmatrix}
    \quad\leftrightarrow\quad
    \ell^{I\mathrlap{^\prime}} \,=\, (U^{-1})^{I\mathrlap{^\prime}}{}_J\mem \ell^J
    \,,
\end{align}
while the twists are related as
\begin{align}
    \label{matr-a}
    \Big(\,{
        \a_{1'}
        \,\,\,
        \a_{2'}
        \,\,\,
        \a_{3'}
    }\,\Big)
    \,\,=\,\,
    \Big(\,{
        \a_{1}
        \,\,\,
        \a_{2}
        \,\,\,
        \a_{3}
    }\,\Big)
    \nem
    \left(\,\,
    \begin{matrix}
        \,\mathclap{1}\,
        \:\:&\:\:
        \,\mathclap{0}\,
        \:\:&\:\:
        \,\mathclap{1}\,
        \\
        \,\mathclap{0}\,
        \:\:&\:\:
        \,\mathclap{1}\,
        \:\:&\:\:
        \,\mathclap{0}\,
        \\
        \,\mathclap{1}\,
        \:\:&\:\:
        \,\mathclap{1}\,
        \:\:&\:\:
        \,\mathclap{0}\,
    \end{matrix}
    \,\,\right)
    \quad\leftrightarrow\quad
    \a_{I'} \,=\, \a_I\mem U^I{}_{I'}
    \,.
\end{align}
It should be clear that the indices explicitly shown here ($1,2,3$ in $\ell^1,\ell^2,\ell^3$ and $\a_1,\a_2,\a_3$, for instance) are homology indices.

The goal is to verify that the respective realisations of the Feynman integral
belong to the same equivalence class in the sense of \eqref{eq:Isim}.

First of all, we evaluate the realisations of the canonical pairing $\langle \a , \ell \rangle$, i.e., the total power:
\begin{subequations}
\label{DX6.P}
\begin{align}
    \label{DX6.Pa}
    P_\text{(a)}
    {}&{}
    \,=\,
        \a_1
        \mdot 
            \ell^1
        +
        \a_2
        \mdot
            \ell^2
        +
        \a_3
        \mdot
            \ell^3
    \,=\,
        \langle \a , \ell \rangle
    \,,\\[0.25\baselineskip]
    \label{DX6.Pb}
    P_\text{(b)}
    {}&{}
    \begin{aligned}[t]
    \,&=\,
    \left(
    \begin{aligned}[c]
        &
        \a_1
        \mdot 
            \ell^1
        +
        \a_2
        \mdot
            \ell^2
        +
        \a_3
        \mdot
            (\ell^1 \mminus p_1)
        +
        \a_3
        \mdot
            (\ell^2 \mplus p_2)
    \\
        &{}
        +
        \a_3
        \mdot
            ( \ell^3 + p_3 -\ell^2 - \ell^1 + p_1 )
    \end{aligned}
    \right)
    \\
    \,&=\,
        \langle \a , \ell \rangle
        + \a_3 \mdot (p_2 \mplus p_3)
    \,,
    \end{aligned}
    \\[0.3\baselineskip]
    \label{DX6.Pc}
    P_\text{(c)}
    {}&{}
    \begin{aligned}[t]
    \,&=\,
        \a_1
        \mdot 
            ( \ell^{1\sprime} + \ell^{3\sprime} )
        +
        \a_2
        \mdot
            \ell^{2\sprime}
        +
        \a_3
        \mdot
            ( \ell^{1\sprime} \mplus \ell^{2\sprime} )
    \,,\\
    \,&=\,
        ( \a_1 \mplus \a_3 ) \mdot \ell^{1\sprime}
        +
        ( \a_2 \mplus \a_3 ) \mdot \ell^{2\sprime}
        +
            \a_1 \mdot \ell^{3\sprime}
    \,,\\
    \,&=\,
        \a_{1'} \mdot \ell^{1\sprime}
        +
        \a_{2'} \mdot \ell^{2\sprime}
        +
            \a_{3'} \mdot \ell^{3\sprime}
    \,=\,
        \langle \a , \ell \rangle
    \,.
    \end{aligned}
\end{align}
\end{subequations}
When deducing the last line in \eqref{DX6.Pc},
we have used \eqref{matr-a}.
Certainly, Eqs.\,(\ref{DX6.Pa})-(\ref{DX6.Pc})
differ only by external factors
and thus
all belong to the same equivalence class:
\begin{align}
    [ P_\text{(a)} ]
    \,=\,
    [ P_\text{(b)} ]
    \,=\,
    [ P_\text{(c)} ]
    \,=\,
        [\mem{ \langle \a , \ell \rangle }\mem]
    \,.
\end{align}

Next, the realisation of the twisted Feynman integral 
which the realisation (a) of the twisted Feynman graph
in \fref{fig:DX6} defines
is
\begin{subequations}
\begin{align}
    \label{DX6.tfi-a}
    I_\text{(a)}
    =
        \int_{\ell^1,\ell^2,\ell^3}
            \frac{
                \mathe^{i \langle \alpha , \ell \rangle}
            }{
                ( \ell^1 )^2
                ( \ell^1 \mminus p_1 )^2
                ( \ell^2 )^2
                ( \ell^2 \mplus p_2 )^2
                ( \ell^3 \mminus \ell^1 \mminus \ell^2 \mplus p_1 \mplus p_3 )^2
                ( \ell^3 \mminus p_4 )^2
                ( \ell^3 )^2
                ( \ell^3 \mplus p_3 )^2
            }
    \,,
\end{align}
which describes a $3d$-dimensional integral
over the loop momentum variables $\ell^1$, $\ell^2$, and $\ell^3$.
Here, we have assumed massless scalar propagators for simplicity.
For the realisation (b) in \fref{fig:DX6}, on the other hand,
the realisation of the twisted Feynman integral is
\begin{align}
    \label{DX6.tfi-b}
    I_\text{(b)}
    \,=\,
        I_\text{(a)}\, \mathe^{i \a_3\cdot(p_2+p_3)}
    \,.
\end{align}
Finally, for the realisation (c) in \fref{fig:DX6},
the realisation of the twisted Feynman integral is
\begin{subequations}
\begin{align}
    \label{DX6.tfi-c*}
    I_\text{(c)}
    \,=
        \int_{\ell^{1\sprime},\ell^{2\sprime},\ell^{3\sprime}}
            \frac{
                \mathe^{i \langle \alpha , \ell \rangle}
            }{
                ( \ell^{1'3'} )^2
                ( \ell^{1'3\sprime} \mminus p_1 )^2
                ( \ell^{2'} )^2
                ( \ell^{2\sprime} \mplus p_2 )^2
                ( \ell^{3\sprime} \mminus p_1 \mminus p_3 )^2
                ( \ell^{1'2\sprime} \mminus p_4 )^2
                ( \ell^{1'2'} )^2
                ( \ell^{1'2\sprime} \mplus p_3 )^2
            }
    \,,
\end{align}
where 
$\ell^{1'2\sprime} :=
    \ell^{1\sprime} \mplus \ell^{2\sprime}
$ and $\ell^{1'3\sprime} :=
    \ell^{1\sprime} \mplus \ell^{3\sprime}
$.
Upon change of variables 
via \eqref{matr-l},
\eqref{DX6.tfi-c*} boils down to\begin{align}
    \label{DX6.tfi-c}
    I_\text{(c)}
    \,=\,
        I_\text{(a)}
    \,.
\end{align}
\end{subequations}
\end{subequations}
Therefore, from \eqrefs{DX6.tfi-b}{DX6.tfi-c},
we conclude that
\begin{align}
    \label{DX6.tfi=}
    [ I_\text{(a)} ]
    \,=\,
    [ I_\text{(b)} ]
    \,=\,
    [I_\text{(c)} ]
    \,=\,
        [ I_\a(\G) ]
    \,.
\end{align}

Note that the conclusion in \eqref{DX6.tfi=}
does not change
when the loop momenta are shifted.
In particular, suppose one performs a shift $\ell^{3'} \mapsto \ell^{3\sprime} \mplus p_1 \mplus p_3$
in the realisation (c).
Then the realisation of the twisted Feynman integral changes to
\begin{align}
    I_\text{(c)}
    \,=\,
        I_\text{(a)}\,
            \mathe^{-i\a_3\cdot(p_1 + p_3)}
    \,.
\end{align}
Still, this is $I_\text{(a)}$ times an external factor.

\section{Representations} \label{sec:REP}

Although the geometric formulation of gauge theories using the language of fibre bundles has greatly advanced our understanding of them, in practical applications\textemdash such as QCD computations for collider observables\textemdash it is often preferred to obscure the geometric structures in favour of facilitating algebraic manipulation, i.e. to fix a gauge. The same is true for twisted Feynman integrals; it is often preferred to fix a realisation when evaluating twisted Feynman integrals for applications.

Several tools and approaches have been developed in the literature for evaluating Feynman integrals. 
We revisit them and generalise them for the study of twisted Feynman integrals, and apply them to the (two-loop) twisted Feynman integral evaluated in \rcite{Kim:2024grz}.
We find that the small deformation by an exponential factor leads to unexpected and novel mathematical properties of twisted Feynman integrals. 
For example, 
the Symanzik polynomials are no longer homogeneous, the (generalised) Baikov parametrisation~\cite{Baikov:1996iu}\textemdash a standard tool for studying the function space of Feynman integrals\textemdash \emph{fails} to determine the function space of twisted Feynman integrals, and exponential periods~\cite{Kontsevich2001}\textemdash rather than periods that describe Feynman integrals~\cite{10.1155/S107379280313142X,Bogner:2007mn,brown2010periodsfeynmanintegrals}\textemdash seem to describe the class of functions that can appear in twisted Feynman integrals.

\subsection{Graph Polynomials}
Consider a $D$-dimensional $L$-loop \emph{Euclidean} integral with $n$ propagators, deformed by the exponential factor $e^{i \sum_{I} \a_I \cdot \ell^I}$. The propagators $\mathcal{D}_e (\ell^I)$ are assumed to be quadratic or linearised(eikonalised) with the following choice of $i0^+$ prescriptions. 
\begin{align}
    I = \int d^{DL}\ell\,\, \frac{e^{i \sum_{I} \a_I \cdot \ell^I}}{\prod_{e=1}^n [\mathcal{D}_e (\ell^I)]^{\n_e}} \,,\quad \mathcal{D}_e = \left\{ 
    \begin{aligned}
        & q_e^2 + m_e^2 && & e &\in E_2
        \\ & q_e \cdot u_e + \omega_e - i 0^+ && & e &\in E_1
    \end{aligned}
    \right.
\end{align}
We assume all vectors (external momenta, $u_e$, and $\a_I$) and scalars ($m_e$ and $\omega_e$) appearing in the integrand to be real-valued, and absorb all numerical factors into the normalisation factor. The quadratic propagators ($e \in E_2$) are positive definite but the linearised propagators ($e \in E_1$) are not positive definite,      
thus for quadratic propagators we use the (Euclidean) Schwinger parametrisation
\begin{align}
    \frac{1}{x^\n} = \frac{1}{\Gamma(\n)} \int_0^\infty dt \, t^{\n - 1} e^{-xt} \,,
\end{align}
while for linearised propagators we use the (Lorentzian) Schwinger parametrisation
\begin{align}
    \frac{1}{(x - i0^+)^\n} = \frac{i^\n}{\Gamma(\n)} \int_0^\infty dt \, t^{\n - 1} e^{-i (x - i0^+)t} \,. \label{eq:STparam_Lor}
\end{align}

After applying the Schwinger parametrisations, the integral takes the form
\begin{align}
\begin{aligned}
    I &= N (\n_e) \int_0^\infty \prod_e \frac{dt_e}{t_e} \, t_e^{\n_e} \int d^{DL}\ell\,\,  \exp \left[ - \sum_{e \in E_2} t_e \mathcal{D}_e - i \sum_{e \in E_1} t_e \mathcal{D}_e + i \sum_I \a_I \cdot \ell^I \right]
    \\ &= N (\n_e) \int_0^\infty \prod_e \frac{dt_e}{t_e} \, t_e^{\n_e} \int d^{DL}\ell\,\, \exp \left[ - M_{IJ} (t_e) (\ell^I \cdot \ell^J) - 2 Q_I (t_e) \cdot \ell^I - R (t_e) + i \a_I \cdot \ell^I \right] \,,
\end{aligned} \label{eq:GP_deriv1}
\end{align}
where $N(\n_e)$ is the normalisation factor irrelevant for studying the graph polynomials, $M_{IJ} (t_e)$ is a symmetric $L \times L$ matrix, $Q_I (t_e)$ is a collection of $L$ vectors, and summation over repeated loop momentum indices ($I$ and $J$) is understood. Note that the $t_e$-dependent part of the exponent is linear in $t_e$, and that the real part of the exponent is negative definite for any $t_e$ due to positive definiteness of quadratic propagators $D_{e \in E_2}$, which makes the gaussian loop integral convergent.

Completing the exponent into sum over squares by introducing the inverse matrix $\tilde{M}^{IJ}$ (defined by the relation $\tilde{M}^{IJ} M_{JK} = \delta^I_K$), the exponent becomes
\begin{align}
\begin{aligned}
    & \exp \left[ - M_{IJ} (t_e) (\ell^I \cdot \ell^J) - 2 Q_I (t_e) \cdot \ell^I - R (t_e) + i \a_I \cdot \ell^I \right]
    \\ &= \exp \left[ - M_{IJ} \left( \ell^I + \tilde{M}^{IK} \left[ Q_K - \frac{i \a_K}{2} \right] \right) \cdot \left( \ell^J + \tilde{M}^{JL} \left[ Q_L - \frac{i \a_L}{2} \right] \right) \right]
    \\ &\phantom{=asdfasd} \times \exp \left[ - \left( R - \tilde{M}^{IJ} \left[ Q_I - \frac{i \a_I}{2} \right] \cdot \left[ Q_J - \frac{i \a_J}{2} \right] \right) \right] \,.
\end{aligned}
\end{align}
Evaluating the gaussian loop integrals, we are left with 
\begin{align}
    I &= \tilde{N} (\n_e, D) \int_0^\infty \prod_e \frac{dt_e}{t_e} \, t_e^{\n_e} \frac{e^{-\frac{\F (t_e)}{\U (t_e)}}}{[\U (t_e)]^{\frac{D}{2}}} \,,
    \\ \U (t_e) &:= \det \left[ M_{IJ} (t_e) \right] \,,
    \\ \F (t_e) &:= \U (t_e) \times \left( R (t_e) - \tilde{M}^{IJ} (t_e) \left[ Q_I (t_e) - \frac{i \a_I}{2} \right] \cdot \left[ Q_J (t_e) - \frac{i \a_J}{2} \right] \right) \,,
    \label{eq:schwinger}
\end{align}
where $\U (t_e)$ and $\F (t_e)$ are the (generalised) first and second Symanzik polynomials respectively, which are collectively referred to as graph polynomials. The negative definiteness of the (real part of the) exponent in \eqref{eq:GP_deriv1} translates to the inequality $\U (t_e) \ge 0$.

While $\U (t_e)$ is a homogeneous polynomial of degree $L$ in Schwinger parameters $t_e$'s, $\F (t_e)$ is \emph{inhomogeneous} and splits into three homogeneous polynomials $\F_{0,1,2} (t_e)$;
\begin{align}
    \U (\l t_e) &= \l^L \, \U (t_e) \,,
    \\ \F (t_e) &= \F_0 (t_e) + \F_1 (t_e) + \F_2 (t_e) \,,\quad \F_i ( \l t_e) = \l^{L + 1 - i} \F_i (t_e) \,.
\end{align}
We define $i$ as the \emph{grade} of the second Symanzik polynomial. The grade of $\F_i (t_e)$ denotes the order of dependence on $\a_I$ vectors;
\begin{align}
\begin{aligned}
    \F_0 (t_e) &:= \U (t_e) \left( R (t_e) - \tilde{M}^{IJ} (t_e) \left[ Q_I (t_e) \cdot Q_J (t_e) \right] \right) \,,
    \\ \F_1 (t_e) &:= +i \,\U (t_e) \left( \tilde{M}^{IJ} (t_e) \left[ Q_I (t_e) \cdot \a_J \right] \right) \,,
    \\ \F_2 (t_e) &:= \frac{1}{4} \U (t_e) \left[ \tilde{M}^{IJ} (t_e) \left( \a_I \cdot \a_J \right) \right] \,.
\end{aligned}
\end{align}
The negative definiteness of the (real part of the) exponent in \eqref{eq:GP_deriv1} translates to the inequalities $\Re [\F_0(t_e)] \ge 0$ (from the condition evaluated at $\a_I = 0$) and $\F_2 (t_2) \ge 0$ (from positive definiteness of the matrix $M_{IJ}$). 

To convert the integral into Feynman parametrisation, we insert decomposition of the identity
\begin{align}
\begin{aligned}
    I &= \tilde{N} (\n_e,D) \int_0^\infty \prod_e \frac{dt_e}{t_e} \, t_e^{\n_e} \frac{e^{-\frac{\F (t_e)}{\U (t_e)}}}{[\U (t_e)]^{\frac{D}{2}}}
    \\ &= \tilde{N} (\n_e,D) \int_0^\infty dt \int_0^\infty \left[ \prod_e \frac{dt_e}{t_e} \, t_e^{\n_e} \right] \delta \left( t - \sum_e t_e\right) \frac{e^{-\frac{\F (t_e)}{\U (t_e)}}}{[\U (t_e)]^{\frac{D}{2}}} \,.
\end{aligned}
\end{align}
Rescaling the Schwinger parameters $t_e$ by $t$ as $t_e = t \times \mathrm{a}_e$ and reversing the order of integration, the integral becomes
\begin{align}
\begin{aligned}
    I &= \tilde{N} (\n_e,D) \int_0^\infty \left[ \prod_e \frac{d \mathrm{a}_e}{\mathrm{a}_e} \mathrm{a}_e^{\n_e} \right] \frac{\delta \left( 1 - \sum_e \mathrm{a}_e\right)}{[\U (\mathrm{a}_e)]^{\frac{D}{2}}} e^{- \P_1 (\mathrm{a}_e)} \int_0^\infty \frac{dt}{t} t^{\n - \frac{LD}{2}} 
    e^{- \P_0 (\mathrm{a}_e) t - \frac{\P_2 (\mathrm{a}_e)}{t}} \,,
    \\ &\P_i (\mathrm{a}_e) := \frac{\F_i (\mathrm{a}_e)}{\U (\mathrm{a}_e)} \,,
\end{aligned}
\end{align}
where $\n = \sum_e \n_e$ and $\mathrm{a}_e$ are Feynman parameters. The $dt$ integral can be rewritten as
\begin{align}
\begin{aligned}
    \int_0^\infty \frac{dt}{t} t^{\n - \frac{LD}{2}} e^{- \P_0 t - \frac{\P_2}{t}} &= \left( \frac{\P_2}{\P_0} \right)^{\frac{2\n - LD}{4}} \int_0^\infty \frac{dt}{t} \, \tilde{t}^{\n - \frac{LD}{2}} e^{- \left( \P_0 \P_2 \right)^{1/2} (\tilde{t} + \tilde{t}^{-1})}
    \\ &= \left( \frac{\F_2}{\F_0} \right)^{\frac{2\n - LD}{4}} \int_0^\infty \frac{dt}{t} \, \tilde{t}^{\n - \frac{LD}{2}} e^{- \frac{\left( \F_0 \F_2 \right)^{1/2}}{\U} (\tilde{t} + \tilde{t}^{-1})} \,,
\end{aligned}
\end{align}
where
\begin{align}
    \tilde{t} &= \left( \frac{\P_0}{\P_2} \right)^{\frac{1}{2}} t = \left( \frac{\F_0}{\F_2} \right)^{\frac{1}{2}} t \,.
\end{align}
The condition $\Re[\F_0] \ge 0$ guarantees that the $dt$ integration contour can be deformed and evaluated analytically using the integral representation of Bessel functions,
\begin{align}
    \int_0^\infty \frac{dt}{t} \, t^{\n} e^{- z (t + {t}^{-1})} = \int_{-\infty}^{+\infty} d\t \, e^{- 2z \cosh{\t}} e^{\n\t} = 2 \int_0^\infty d\t e^{-2z \cosh{\t}} \cosh (\n \t) = 2 K_{\n} (2z) \,, \label{eq:BFKintrep}
\end{align}
which holds for $|\arg z| < \frac{\pi}{2}$; see Eq.(10.32.9) of DLMF~\cite{dlmf}. 
The appearance of Bessel functions is a special feature of twisting; we only get rational functions of Feynman parameters $\mathrm{a}_e$ raised to some powers from the $dt$ integral for usual Feynman integrals. 
Collecting all ingredients, the integral becomes
\begin{align}
\begin{aligned}
    I &= N' (\n_e,D) \int_0^\infty \left[ \prod_e \frac{d \mathrm{a}_e}{\mathrm{a}_e} \mathrm{a}_e^{\n_e} \right] \frac{\delta \left( 1 - \sum_e \mathrm{a}_e\right)}{[\U (\mathrm{a}_e)]^{\frac{D}{2}}} \left( \frac{\F_2 (\mathrm{a}_e)}{\F_0 (\mathrm{a}_e)} \right)^{\frac{2\n - LD}{4}} e^{- \P_1 (\mathrm{a}_e)} K_{\n - \frac{LD}{2}} \left( 2 \tilde{\P} (\mathrm{a}_e) \right)
    \\ &= N' (\n_e,D) \int_0^\infty \left[ \prod_e \frac{d \mathrm{a}_e}{\mathrm{a}_e} \mathrm{a}_e^{\n_e} \right] \delta \left( 1 - \sum_e \mathrm{a}_e\right) \frac{[\U(\mathrm{a}_e)]^{\n - \frac{(L+1)D}{2}}}{[\F_0 (\mathrm{a}_e)]^{\n - \frac{LD}{2}}}
    \\ &\phantom{=asdfasdfasdfasdf} \times \left\{ [\tilde{\P} (\mathrm{a}_e)]^{\n - \frac{LD}{2}} K_{\n - \frac{LD}{2}} \left( 2 \tilde{\P} (\mathrm{a}_e) \right) e^{- \P_1 (\mathrm{a}_e)} \right\} \,,
\end{aligned} \label{eq:TFI_FP}
\end{align}
where
\begin{align}
    \tilde{\P} (\mathrm{a}_e) &:= \frac{\left( \F_0 (\mathrm{a}_e) \F_2  (\mathrm{a}_e) \right)^{1/2} }{\U (\mathrm{a}_e)} \,.
\end{align}
The curly brackets in the second line reduces to $\frac{1}{2} \Gamma (\n - \frac{LD}{2})$ in the limit $\a_I \to 0$, recovering the usual form of the graph polynomials in Feynman parametrisation.

\subsection{Baikov Parametrisation} \label{sec:BP}

The Baikov parametrisation~\cite{Baikov:1996iu} is another parametrisation of Feynman integrals widely used to investigate their underlying geometry. 
The inverse propagators are used as integration variables in this parametrisation, 
and ``cutting'' a propagator $x_j$ corresponds to taking the residue of the integrand at $x_j = 0$.

There are two types of Baikov parametrisations.
The first is the standard Baikov parametrisation. 
The other is the loop-by-loop Baikov parametrisation, which is applying the standard Baikov parametrisation to each loop at a time. 
The latter has the advantage that often fewer variables are needed than the standard parametrisation, due to less irreducible scalar products (ISPs) appearing in the sub-loop integrals.\footnote{The loop-by-loop Baikov parametrisation is not unique and depends on the order of loop momenta converted to the Baikov parametrisation.} 
For a Feynman integral in momentum space
\begin{align}
    I= \int d^{DL}\ell\;\frac{ N\left(\ell\right) \, }{\prod_{i=1}^P \mathcal{D}_i^{a_i}},
\end{align}
with $P$ number of propagators $\mathcal{D}_i$ and a generic scalar numerator $N\left(\ell\right)$,
the loop-by-loop Baikov parametrisation is given as
\begin{align}
    I=\frac{\mathcal{J}\pi^{\frac{(D+1)L-n}{2}}}{\prod_{l=1}^L\Gamma\left(\frac{D-E_l}{2}\right)}\int_{\mathbf{C}_{B}}d^nx\frac{x^{-a_{P+1}}_{P+1}\cdots x_n^{-a_n}}{x_1^{a_1}\cdots x_{P}^{a_P}}\left(\prod_{l=1}^L \mathcal{E}^{\frac{E_l-D+1}{2}}_l\mathcal{B}_l^{\frac{D-E_l-2}{2}}\right) \,, \label{eq:LBL_Baikov_ansatz}
\end{align}
where $\mathcal{J}$ is the Jacobian factor, $L$ is the number of loops, $P$ is the number of propagators, $E_l$ is the number of independent external momenta of the $l$-th loop and the number of integration variables is $n=L+\sum_{l=1}^L E_l$.
Each loop has two polynomials given by
\begin{align}
    \mathcal{E}_l=\det\left(G\left(q_1,\dots,q_{E_l}\right)\right)\,,\qquad \mathcal{B}_l=\det \left(G\left(\ell_l,q_1,\dots,q_{E_l}\right)\right)\,, \label{eq:EBPolyDef}
\end{align}
where the Gram matrix $G\left(\{q\}\right)$ is the matrix constructed by all the scalar products $G_{ij}=q_i\cdot q_j$, $q_l$ are the external momenta of $l$th loop and $\ell_l$ is the internal momentum of the loop.
The integration domain $\mathbf{C}_B$ is 
\begin{align}
    \mathbf{C}_B=\Biggl\{x\in \mathbb{R}:\bigwedge_{l=1}^L \frac{\mathcal{B}_l\left(x\right)}{\mathcal{E}_l\left(x\right)}>0\Biggl\} \,.
\end{align}
Note that \eqref{eq:LBL_Baikov_ansatz} can be considered as the integral of a multivariate algebraic function within a domain specified by an algebraic function, which is the definition of a (algebraic) \emph{period}; Feynman integrals can be considered as periods~\cite{10.1155/S107379280313142X,Bogner:2007mn,brown2010periodsfeynmanintegrals}. 

We generalise the Baikov parametrisation to twisted Feynman integrals following the approach of \rcite{Brunello:2023fef}.
Consider a twisted Feynman integral given as
\begin{align}
    I= \int d^{DL}\ell\; \exp\left(\sum_{j}\mathcal{D}_j\right)\frac{N\left(\ell\right) }{\prod_{i=1}^P \mathcal{D}_i^{a_i}} \,.
\end{align}
Its loop-by-loop Baikov parametrisation reads
\begin{align}
    I=\frac{\mathcal{J}\pi^{\frac{(D+1)L-n}{2}}}{\prod_{l=1}^L\Gamma\left(\frac{D-E_l}{2}\right)}\int_{\mathbf{C}_B}d^nx \exp\left(\sum_{j}x_{j}\right)\frac{x^{-a_{P+1}}_{P+1}\cdots x_n^{-a_n}}{x_1^{a_1}\cdots x_{P}^{a_P}}\left(\prod_{l=1}^L \mathcal{E}^{\frac{E_l-D+1}{2}}_l\mathcal{B}_l^{\frac{D-E_l-2}{2}}\right) \,,
\end{align}
where the definitions for the integration variables $x_j$, polynomials $\mathcal{E}_l$ and $\mathcal{B}_l$, and the integration domain $\mathbf{C}_B$ remain unchanged. Note that the difference from the usual Feynman integral [\eqref{eq:LBL_Baikov_ansatz}] is the additional exponential factor, which implies that twisted Feynman integrals are given by \emph{exponential periods} rather than periods. 
We will indeed see that some twisted Feynman integrals evaluate to (sums over) values of Bessel functions in sec.~\ref{sec:TFI_DC}; the values of Bessel functions are known to be exponential periods~\cite{Kontsevich2001}.

\subsubsection*{Revealing geometry through leading singularity}

For special kinematic configurations, Feynman integrals are known to develop singularities associated with physical processes such as particle production. 
Such singularities can be studied through the Landau equations~\cite{Landau:1959fi}, where the solutions to the Landau equations\textemdash the Landau singularities\textemdash provide the possible singular loci of the Feynman integral. 
Moreover, a subset of Landau singularities appears as kernels of integration when Feynman integrals are evaluated using the method of differential equations; see sec.~\ref{sec:DE}. 
Therefore, the function space and the underlying geometry of a Feynman integral can be explored by studying its Landau singularities.

It is known that the leading singularity of a one-loop Feynman integral (the maximal residues of its integrand) corresponds to the inverse square root of the (leading) Landau singularity, and the correspondence can be extended to multi-loop integrals through a loop-by-loop analysis~\cite{Flieger:2022xyq}. 
Therefore, the Landau singularities of a Feynman integral can be understood from its leading singularities. 
The Baikov parametrisation provides a convenient setup to study the leading singularities of a Feynman integral, which in turn provides the information of the geometry underlying it; see, e.g., \rcite{Brammer:2025rqo}. 
Given the generalisation of the Baikov parametrisation to twisted Feynman integrals, the same approach can be used to study the underlying geometry of twisted Feynman integrals. 
We show that this approach, in fact, fails.

The leading singularity of a Feynman integral $I$ takes the form
\begin{align}
    \text{LS}\left(I\right)\propto \int\frac{d^n\Vec{z}}{Q \left(\Vec{z}\right)} \,.
\end{align}
The form of $Q(\Vec{z})$ determines the function space of the Feynman integral: when the denominator is a polynomial of degree $m$, $Q(\Vec{z}) = P_m (\Vec{z})$, or can be rationalised, $Q(\Vec{z}) = \sqrt{[P_m (\Vec{z})]^2}$, the integral can be evaluated and the resulting leading singularity is an algebraic function. 
In this case the function space is given by polylogarithmic functions, which are functions given as iterated integrals of $d\log$s.
On the other hand, if the denominator $Q(\Vec{z}) = \sqrt{P_m (\Vec{z})}$ cannot be rationalised with $m = 2(n+1)$, then the integral family of $I$ is given as (iterated) integrals over an $n$-dimensional Calabi-Yau manifold, where $n = 1$ (an elliptic curve or a torus) is the special case of elliptic integrals.

We show that this argument does not apply to twisted Feynman integrals. 
In particular, we apply the procedure to the example computed in \rcite{Chen:2024bpf}, which is known to involve elliptic integrals. 
The integral family is defined as 
\begin{gather}
    I_{n_1 \cdots n_9}[a , b] = \int_{q,\ell} \frac{ e^{\mathcal{D}_1 + \mathcal{D}_2} \, \delta(\mathcal{D}_3) }{\prod_{i=1}^9 \mathcal{D}_i^{n_i}} \,, \label{eq:ibpset} \\
    \begin{aligned}
        \mathcal{D}_1 &=  ib \cdot q \,,\quad & \mathcal{D}_2 &= a \cdot \ell \,,\quad & \mathcal{D}_3 &= v \cdot q \,,
        \\ \mathcal{D}_4 &= \ell^2 \,,\quad & \mathcal{D}_5 &= (q-\ell)^2 \,,\quad & 
        \mathcal{D}_6 & = v \cdot \ell \,,
        \\ \mathcal{D}_7 &= b \cdot \ell \,,\quad  & \mathcal{D}_8 &= a \cdot q \,,\quad  & \mathcal{D}_9 &= q^2 \,,
    \end{aligned}
    \label{eq:2L_ISPs}
\end{gather}
where $\mathcal{D}_i$ with $i>6$ are additional ISPs required to close the system. We set $n_i\leq 0$ for $i>6$. 
The simplest nontrivial integral in this family can be represented as $I_{0,0,1,1,1,1,0,0,0}$; its loop-by-loop Baikov parametrisation reads
\begin{gather}
    \begin{aligned}
            I_{0,0,1,1,1,1,0,0}\propto \int\prod_{i=1}^{8}&d x_i e^{\left(x_1+x_2\right)}\left.\frac{1}{x_4 x_5 x_6}\left(\prod_{l=1}^2 \mathcal{E}_l^{\left(E_l-d+1\right)/2}+\mathcal{B}_l^{\left(d-E_l-2\right)/2}\right)\right|_{x_3=0} \,,
    \end{aligned}
\end{gather}
where $\mathcal{E}_l$ and $\mathcal{B}_l$ are the polynomials defined in \eqref{eq:EBPolyDef}. 
For the considered example, the simplification of the loop-by-loop Baikov parametrisation occurs and only 8 variables $x_1,\dots,x_8$ appear in the expression, which corresponds to the ISPs $\mathcal{D}_i$ excluding $\mathcal{D}_7$. 
The maximal cut is given as 
\begin{gather}
    \begin{aligned}
        &\left.I_{0,0,1,1,1,1,0,0}\right|_{\text{max.cut}}
        \propto\int d x_1 dx_2 dx_7 dx_8 e^{x_1+x_2} \frac{\sqrt{-4 m_v x_7^2+4 m_a m_v x_8 -s_{13}^2 x_8}}{\mathcal{Q}} \,,
    \end{aligned} \label{eq:TFI_LS}
\end{gather}
where 
\begin{gather}
    \begin{aligned}
        \mathcal{Q} &= x_8 \left(16m_vx_2^2-16m_vx_2x_7+4m_am_vx_8-s_{13}^2x_8 \right)
        \\ &\phantom{=as} \times\left(4m_am_v x_1^2-s_{13}^2x_1^2-4 i m_vs_{12}x_1x_7-4m_bm_vx_7^2 
        \right. \\ &\phantom{=as} \left. \phantom{asdf} 
        +4m_am_bm_vx_8-m_vs_{12}^2x_8-m_b s_{13}^2x_8\right) \,,
    \end{aligned}
\end{gather}
with the Lorentz invariants $\vec{s}=\{m_a,m_b,m_v,s_{12},s_{13}\}$ (See Section~\ref{sec:DE} for the explicit definition). 
The leading singularities are obtained by taking the residues around the contours of $\mathcal{Q}=0$, and they can be checked to be always algebraic.\footnote{The exponential factors always cancel when taking the maximal residues, which leads to an algebraic expression.} 
This is also true for the subsector integrals after a change of the variables for rationalisation, 
which implies that the resulting integrals are of polylogarithmic form. 
However, the correct function space is spanned by more complicated integrals than polylogarithms, which can be determined by the method of differential equations~\cite{Chen:2024bpf}. 
We conclude that the leading singularities of twisted Feynman integrals do not faithfully represent their function space, and additional procedures seem to be necessary for studying the underlying geometry of twisted Feynman integrals through leading singularities.

Ref.~\cite{Chen:2024bpf} introduced an approach where the integrand of twisted Feynman integrals can be converted to rational functions without exponential factors, so that the integrals can be approached using established multi-loop evaluation techniques without modifications. The price one pays is an extra Fourier transform,
\begin{gather}
        I[a,b]=\int dt e^{it}\Tilde{I}[t;a,b] \,, \label{eq:FTedTFI}
\end{gather}
where $t$ is an auxiliary parameter for the Fourier transform and $\tilde{I}\left[t;a,b\right]$ is defined as
\begin{gather}
    \tilde{I}_{n_1 \cdots n_9}\left[t;a,b\right]=\int_{q,l}\frac{\delta\left(\mathcal{D}_1+\mathcal{D}_2-it\right)\delta\left(\mathcal{D}_3\right)}{\prod_{i=1}^9 \mathcal{D}_i^{n_i}} \,. \label{eq:FTedTFI2}
\end{gather}
Exchanging the order of integration between $dt$ and loop momenta in \eqref{eq:FTedTFI}, we recover the original integrand \eqref{eq:ibpset}.

In the loop-by-loop Baikov parametrisation, the integral of \eqref{eq:FTedTFI2} becomes
\begin{gather}
    \begin{aligned}
        \Tilde{I}_{0,0,1,1,1,1,0,0}\propto \int dx_1 dx_4 dx_5 dx_6 dx_7 dx_8 \left.\frac{1}{x_4 x_5 x_6}\left(\prod_{l=1}^2 \mathcal{E}_l^{\left(E_l-d+1\right)/2}+\mathcal{B}_l^{\left(d-E_l-2\right)/2}\right)\right|_{\substack{x_3=0\\x_2=it-x_1}} \,,
    \end{aligned}
\end{gather}
where delta constraints have been removed by the $dx_2 dx_3$ integrals. The maximal residues of this integral also turn out to be algebraic, which take the form
\begin{gather}
    \left.\tilde{I}_{0,0,1,1,1,1,0,0}\right|_{\text{max.cut}}=\frac{\mathcal{N
    }\left(\vec{s}\right)}{t},
\end{gather}
where $\mathcal{N}(\vec{s})$ is an algebraic function. Consequently, the leading singularities of the original integral are obtained by taking the residue further at $t=0$,
\begin{gather}
    \left.I_{0,0,1,1,1,1,0,0}\right|_{\text{max.cut}}= \oint_{\mathcal{C}_t} dt\,\mathcal{N}\left(\vec{s}\right)\frac{e^{it}}{t}\propto \mathcal{N}\left(\vec{s}\right), \label{eq:FTedLS}
\end{gather}
which are algebraic expressions.\footnote{For subsector integrals, the remaining $dt$ integral takes the form $\int dt \,\mathcal{N} (\vec{s}) \frac{e^{it}}{{t}^n}$ where $\mathcal{N}(\vec{s})$ is an algebraic function and $n$ can be a number other than 1, such as $\frac{5}{2}$ or 3. Although the dependence on $t$ changes, the integrand maintains the factorisable $\propto \frac{e^{it}}{t^n}$ form, therefore the leading singularity is proportional to $\mathcal{N}(\vec{s})$ and remains algebraic.}

\subsubsection*{Checking the number of critical points}

The integrals given in App.~B of \rcite{Kim:2024grz} provide concrete examples of evaluated twisted Feynman integrals, which can serve as nontrivial examples to test the conclusions of our analyses. The explicit evaluation of the examples will be provided in sec.~\ref{sec:TFI_DC}.

The number of independent functions\textemdash such as the master integrals; see sec.~\ref{sec:DE}\textemdash that span the function space of the integral family can be determined using the Baikov parametrisation~\cite{Lee:2013hzt}.
As an example, we consider the integral computed in \rcite{Chen:2024bpf},
\begin{align}
    I [a,b] &:= \int d^D q \, d^D \ell \, \frac{e^{i b \cdot q} \,  e^{a \cdot \ell} \, \delta(v\cdot q)}{(\ell^2)^{\lambda_1}[(q - \ell)^2]^{\lambda_2} (v \cdot \ell - i0^+)^{\lambda_3}} \,,
    \label{eq:integral_check_MIs}
\end{align}
which corresponds to the integral $I_{0,0,1,\l_1,\l_2,\l_3,0,0,0}$ in the notations of \eqref{eq:ibpset}, with Wick rotation $b \to ib$. Our example integral take the form of
\begin{align}
    I=\int_{\mathbf{C}_R}u\left(x_e\right)\varphi\left(x_e\right) \,,
\end{align}
where 
\begin{gather}
    \begin{aligned}
        u\left(x_e\right)&=e^{\left(x_1+x_2\right)}\left.\frac{1}{x_4^{\nu_4} x_5^{\nu_5} x_6^{\nu_6}}\left(\prod_{l=1}^2 \mathcal{E}_l^{\left(E_l-d+1\right)/2}+\mathcal{B}_l^{\left(d-E_l-2\right)/2}\right)\right|_{x_3=0}\,,\\
        \varphi\left(x_e\right)&=\prod_{e}\frac{dx_e}{x_e} \,.
    \end{aligned}
\end{gather}
Critical points are defined as solutions to the equation $d\log u\left(x_e\right)=0$, and the number of critical points is equal to the number of master integrals. 
We find 4 critical points, which we confirm in sec.~\ref{sec:DE}.

\subsection{Differential Equations} \label{sec:DE}

It is often very difficult to directly evaluate Feynman integrals. However, evaluating the integrals in special kinematic limits is often possible, and we can determine the desired Feynman integrals from the integrals evaluated in special kinematics through the relations between the integrals. This is the key idea behind the method of differential equations~\cite{Kotikov:1990kg,Kotikov:1991hm,Kotikov:1991pm,Gehrmann:1999as}, and the idea can also be extended to the determination of twisted Feynman integrals~\cite{Brunello:2024ibk,Hu:2024kch,Chen:2024bpf,Chen:2025gqu,Feng:2025leo}. 

The method of differential equations is founded on the observation that various Feynman integrals satisfy linear relations known as integration-by-parts (IBP) relations, which is a differential statement of translation invariance.\footnote{This means IBP relations should be used with care when the axiom of translation invariance is violated, notably in chiral theories where (ABJ-type) anomalies can become relevant.} Schematically denoting the set of loop momenta by the vector $\vec{\ell}$, the IBP relations can be stated as
\begin{align}
    \int_{\vec{\ell}} \, f (\vec{\ell}) = \int_{\vec{\ell}} \, f (\vec{\ell} + \vec{p}) \quad \Rightarrow \quad \int_{\vec{\ell}} \, \lim_{\epsilon \to 0} \frac{f (\vec{\ell} + \epsilon \vec{p}) - f (\vec{\ell})}{\epsilon} = \vec{p} \cdot \left( \int_{\vec{\ell}} \nabla_{\vec{\ell}} [f (\vec{\ell})] \right) = 0 \,,
\end{align}
where $f(\vec{\ell})$ stands for any integrand and $\vec{p}$ can be any set of constant vectors. An important consequence of IBP relations is that only a finite number of Feynman integrals\textemdash called master integrals (MIs)\textemdash are linearly independent within a family of Feynman integrals called sectors, and all other Feynman integrals within the same sector can be expressed using the master integrals~\cite{Smirnov:2010hn,Bitoun:2017nre}. 

In the method of differential equations, we consider the master integrals of a given sector as a function of external kinematic variables. When taking a derivative with respect to external variables, the master integrals become new integrands which can be re-expressed as linear combinations of master integrals through the IBP relations. 
The desired Feynman integrals are then obtained by (iteratively) integrating the differential equations, where the integrals at special kinematics that can be directly evaluated serve as the initial conditions. 
In this approach, Feynman integrals are expressed as integrals over algebraic varieties that are determined by the integration kernels; the kernels determine the underlying geometry of the Feynman integrals. 

When the integration kernel consists only of rational functions, 
the kernel can be decomposed into logarithmic one-forms ($d\log$ forms) which are called the \emph{symbol letters}, 
and the underlying geometry is the Riemann sphere. 
The set of symbol letters is called the \emph{(symbol) alphabet}, 
and the Feynman integral is evaluated as iterated integrals of the alphabet, 
which result in polylogarithmic functions; see sec.~3 of \rcite{goncharov2013simpleconstructiongrassmannianpolylogarithms}.
On the other hand, if the kernel cannot be decomposed into $d\log$ forms, 
the function space is more complex than polylogarithms and the Feynman integral must be written as integrals over more complicated geometries such as elliptic curves (elliptic integrals) or Calabi-Yau manifolds.

As remarked, the method of differential equations generalises to twisted Feynman integrals, and the associated integration kernel determines its underlying geometry. We consider the integral family defined by \eqref{eq:ibpset} and \eqref{eq:2L_ISPs} as a concrete example.
IBP equations are derived from the fact that the integration of the total derivative of the integrand in dimensional regularisation vanishes. 
\begin{align}
    \int_{q,l}\frac{\partial}{\partial\{\ell^{\mu},q^{\mu}\}}\left(k^{\mu}\cdot\frac{e^{\mathcal{D}_1+\mathcal{D}_2} \; \delta (\mathcal{D}_3)}{\prod_{i=1}^9 \mathcal{D}_i^{n_i}}\right)=0,
\end{align}
with vector $k$ as any linear combination of the external and internal momenta.
By solving the linear system of IBP equations using \texttt{FiniteFlow}~\cite{Peraro:2019svx}, we can obtain the master integrals as
\begin{align}
    \vec{f}=\{ I_{001111000}, I_{00111-1000},I_{-101110000},I_{001110000}\}.
\end{align}
Now, we need to take the derivatives of the master integrals to get a system of differential equations.
The integrals depend only on the scalar product of the external momenta and we define the set of them as $\vec{s}=$\{$m_a^2=a^2$, $m_b^2=b^2$, $m_v^2=v^2$, $s_{12}=\left(a+b\right)^2$, $s_{13}=\left(a+v\right)^2$\}, where we set $(b \cdot v)=0$ imposed by the integrand factor $\delta (\mathcal{D}_3)$.
The derivatives with respect to the variables can be obtained by constructing an ansatz as
\begin{gather}
    \begin{aligned}
        \frac{\partial}{\partial s_i}=&\left(\a^i_1 a^{\m}+ \a^i_2 b^{\m}+\a^i_3 v^{\m}\right)\frac{\partial}{\partial a^{\m}}
        +\left(\b^i_1 a^{\m}+ \b^i_2 b^{\m}+\b^i_3 v^{\m}\right)\frac{\partial}{\partial b^{\m}}+\left(\c^i_1 a^{\m}+ \c^i_2 b^{\m}+\c^i_3 v^{\m}\right)\frac{\partial}{\partial v^{\m}}.
    \end{aligned}
\end{gather}
Each differential operator, for instance $\frac{\partial}{\partial a^{\mu}}$, acts on the integral as
\begin{gather}
    \begin{aligned}
        &\int \frac{\partial}{\partial a^{\mu}}\left(\frac{e^{\mathcal{D}_1+\mathcal{D}_2}\delta\left(\mathcal{D}_3\right)}{\prod_{i=1}^{9}\mathcal{D}_{i}^{n_i}}\right)\\
        &=\int \left(\frac{\partial}{\partial a^{\mu}}e^{\mathcal{D}_1+\mathcal{D}_2}\right)\left(\frac{\delta\left(\mathcal{D}_3\right)}{\prod_{i=1}^{9}\mathcal{D}_{i}^{n_i}}\right)+\int e^{\mathcal{D}_1+\mathcal{D}_2}\left(\frac{\partial}{\partial a^{\mu}}\left(\frac{\delta\left(\mathcal{D}_3\right)}{\prod_{i=1}^{9}\mathcal{D}_{i}^{n_i}}\right)\right).
    \end{aligned}
\end{gather}
The explicit value of each $\alpha$'s, $\beta$'s and $\gamma$'s can be (partially) fixed by how the differential operators act on the scalar products.
For example, the operator $\frac{\partial}{\partial s_{12}}$ should satisfy the relations
\begin{gather}
    \begin{aligned}
        \frac{\partial}{\partial s_{12}}a^2 &=0\,,\quad &
        \frac{\partial}{\partial s_{12}}b^2 &=0\,,\quad
        &\frac{\partial}{\partial s_{12}}v^2 &=0\,,
        \\ \frac{\partial}{\partial s_{12}}(a\cdot b) &=\frac{1}{2}\,,\quad &\frac{\partial}{\partial s_{12}}(b\cdot v) &=0\,,\quad&\frac{\partial}{\partial s_{12}}(a\cdot v)  &=0 \,.
    \end{aligned}
\end{gather}
Since there are six equations while there are nine variables to fix, the remaining three variables can be chosen freely. 
The derivatives of the integrals must not depend on the choice of the remaining three variables, and this property can be used as a consistency check of the derivatives.
For example, applying the differential operator $\frac{\partial}{\partial s_{12}}$ to the first master integral yields
\begin{gather}
    \begin{aligned}
         &\frac{\partial}{\partial s_{12}}I_{001111000}=c_1 I_{0-11111000}+c_2 I_{001110000}+c_3 I_{001111-100},
    \end{aligned} \label{eq:MIder_ex1}
\end{gather}
where $c_i$ are rational functions of $\vec{s}$. 
The right-hand side of \eqref{eq:MIder_ex1} can be reorganised as a linear combination of master integrals through IBP relations, which closes the system of differential equations as linear operators acting on the vector of master integrals $\vec{f}$,
\begin{align}
    \frac{\partial}{\partial s_{i}}\vec{
    f}=A_{s_i}\left(\vec{s};\epsilon\right)\vec{f} \,, \label{eq:MI_DE_system}
\end{align}
where we take $D=3-2\epsilon$.
The differential equations satisfy the following properties.
\begin{itemize}
    \item They satisfy scaling property such as
    $\sum_{i=1}^5 s_i A_{s_i}=\diag \left(2-D,2-D,2-D,2-D\right)$
    \item They satisfy integrability condition as 
    $\partial_{s_i}A_{s_j}-\partial_{s_j}A_{s_i}=\left[A_{s_i},A_{s_j}\right]$ with $i,j=1,\dots,5$
\end{itemize}

\subsubsection*{Revealing geometry through differential equations}
The system of differential equations for the master integrals [\eqref{eq:MI_DE_system}] directly encodes the underlying geometry of their function space, 
since the master integrals are obtained by integrating the differential equations. 
The differential equations can be reorganised as Picard\textendash Fuchs equations by taking further derivatives, e.g., for the master integral $f_k$
\begin{align}
    \mathcal{L}_n^{\left(s_i\right)} f_k=\left(\frac{d^n}{ds_i^n}+\sum_{j=0}^{n-1}c_j\left(\vec{s}\right)\frac{d^j}{ds_i^j}\right)f_k = 0 \,,
\end{align}
where $\mathcal{L}_n$ is called the Picard\textendash Fuchs operator.
The operator $\mathcal{L}_n$ can be factorised into a product of irreducible differential operators $\mathcal{L}_m$ of varying orders $m$. 
If an irreducible operator $\mathcal{L}_m$ with $m>1$ appears in the decomposition, the integral cannot be given as iterated integrals over a Riemann sphere.

When restricting the system of differential equations \eqref{eq:MI_DE_system} to the $(3\times3)$ dimensional subsector spanned by the master integrals $\{ I_{00111-1000},I_{-101110000},I_{001110000}\}$, the system can be reorganised as a Picard\textendash Fuchs equation where the Picard\textendash Fuchs operator $\mathcal{L}_3$ factorises into a second order operator $\mathcal{L}_2$ and a first order operator $\mathcal{L}_1$. 
The presence of an irreducible second order differential operator $\mathcal{L}_2$ indicates that the function space of the family of integrals \eqref{eq:ibpset} cannot be polylogarithmic functions, and the underlying geometry is more complicated than the simple Riemann sphere implied by the algebraic leading singularity \eqref{eq:TFI_LS}.
Note that the situation is not very different when the integrand is reorganised into a rational function through a Fourier transform: although the leading singularities are algebraic as shown in \eqref{eq:FTedLS}, the differential operator nevertheless detects that the underlying geometry is an elliptic curve~\cite{Chen:2024bpf}.

\subsection{Checking against direct computations} \label{sec:TFI_DC}

Consider the following Euclidean one-loop integral with the condition $v \cdot q = v \cdot a = 0$, which is a special case of Eq.~(B.2) evaluated in \rcite{Kim:2024grz}. 
This integral is an infinite sum of modified Bessel functions of the first kind $I_n(z)$.
\begin{align}
\begin{aligned}
    I^{(\l_1, \l_3)} [a,q,v] &:= \int \frac{d^D \ell \; e^{2 a \cdot \ell}}{(\ell^2)^{\lambda_1}[(q - \ell)^2]^{\lambda_1} (2 v \cdot \ell - i0^+)^{\lambda_3}}
    \\ &= \frac{2^{\frac{1-D+2 \l_1-\l_3}{2}} \pi^{1 + \frac{D}{2}} \, i^{\l_3}}{\Gamma(\l_1)^2 \Gamma (\frac{1 + \l_3}{2})} \frac{e^{q \cdot a}}{(v^2)^{\frac{\l_3}{2}} (q^2)^{\frac{4 \l_1 + \l_3 - D}{2}}}
    \\ &\phantom{=asdf} \times \sum_{n=0}^\infty \frac{\Gamma(\a(n)) \Gamma(\l_1 - \a(n))}{n!} \frac{I_{\a(n) - \frac{1}{2}} (q \cdot a)}{(q \cdot a)^{\a(n) - \frac{1}{2}}} \left( \frac{a^2 q^2}{2} \right)^n \,,
    \\ \a(n) &:= \frac{D}{2} - \l_1 - \frac{\l_3}{2} + n \,.
\end{aligned} \label{eq:TFI_ex_1loop}
\end{align}
In particular, the product $\Gamma(\a) \Gamma (\l_1 - \a)$ can be reduced using the reflection formula for $\l_1 \in \mathbb{Z}^{+}$, e.g. for $\l_1 = 1$,
\begin{align}
    I^{(1,0)} [a,q] &= \int \frac{d^D \ell \; e^{2 a \cdot \ell}}{\ell^2 \, (q - \ell)^2} = - \frac{2^{\frac{3 - D}{2}} \pi^{\frac{3 + D}{2}}}{\sin (\frac{\pi D}{2})} \frac{e^{q \cdot a}}{(q^2)^{\frac{4 - D}{2}}} \sum_{n=0}^{\infty} \frac{\left( - \frac{a^2 q^2}{2} \right)^n}{n!} \frac{I_{\frac{D - 3}{2} + n} (q \cdot a)}{(q \cdot a)^{\frac{D - 3}{2} + n}} \,,
    \\ I^{(1,1)} [a,q,v] &= \int \frac{d^D \ell \; e^{2 a \cdot \ell}}{\ell^2 \, (q - \ell)^2 \, (2 v \cdot \ell - i0^+)} \nonumber \\
    &= \frac{2^{1 - \frac{D}{2}} \pi^{2 + \frac{D}{2}} i}{\cos (\frac{\pi D}{2})} \frac{e^{q \cdot a}}{(v^2)^{\frac{1}{2}} (q^2)^{\frac{5 - D}{2}}} \sum_{n=0}^{\infty} \frac{\left( - \frac{a^2 q^2}{2} \right)^n}{n!} \frac{I_{\frac{D}{2}+ n - 2} (q \cdot a)}{(q \cdot a)^{\frac{D}{2}+ n - 2}} \,.
\end{align}
For null $a^\m$ ($a^2 = 0$), the infinite sum localises to $n=0$ and reduces to the modified Bessel function $I_{\frac{D-1}{2} - \l_1 - \frac{\l_3}{2}}(q \cdot a)$.
The appearance of the modified Bessel functions is not very surprising as Bessel functions appear in the Feynman parametrisation [\eqref{eq:TFI_FP}]. 
However, the values of Bessel functions fall under the class of numbers called exponential periods~\cite{Kontsevich2001}, which is a larger class of numbers than periods which the usual Feynman integrals are known to fall under~\cite{10.1155/S107379280313142X,Bogner:2007mn,brown2010periodsfeynmanintegrals}.

Now, consider the effective two-loop integral \eqref{eq:integral_check_MIs}, which was used as the main example for the application of various mathematical tools developed in this section. The integral can be evaluated by composing the integrals given in App.~B of \rcite{Kim:2024grz}:
\begin{align}
\begin{aligned}
    I [a,b] &:= \int d^D q \, d^D \ell \, \frac{e^{i b \cdot q} \,  e^{a \cdot \ell} \, \delta (v \cdot q)}{(\ell^2)^{\lambda_1}[(q - \ell)^2]^{\lambda_2} (v \cdot \ell - i0^+)^{\lambda_3}}
    \\ &= \frac{(4 \pi)^{D - \frac{1}{2}} \, i^{\l_3}}{2^{1 + 2(\l_1 + \l_2)} \Gamma(\l_1) \Gamma (\l_2) \Gamma (\l_3)} \frac{
    \sum_{l,m,n=0}^{\infty} \frac{N_{l,m,n}}{l!\,m!\,n!} \left( 2i \frac{(a \cdot b)}{ b^2 } \right)^{l} \left( -2i \frac{(v \cdot a)}{\sqrt{v^2 b^2}} \right)^{m} \left( \frac{a^2}{b^2} \right)^{n}
    }{(v^2)^{\frac{1 + \l_3}{2}} (b^2)^{D - \frac{1}{2} - \l_1 - \l_2 - \frac{\l_3}{2}}} \,,
    \\ N_{l,m,n} &:= \frac{\Gamma (D - \frac{1}{2} - \l_1 - \l_2 - \frac{\l_3}{2} + l + \frac{m}{2} + n) \Gamma ( \frac{\l_3 + m}{2} )}{\Gamma (D - \l_1 - \l_2 - \l_3 + l + m + 2n)}
    \\ &\phantom{=as} \times \sum_{k=0}^{\infty} {n \choose k} \Gamma (\frac{D - 2\l_1 - \l_3 + 2l + m + 2n + 2k }{2}) \Gamma (\frac{D - 2\l_2 - \l_3 + m + 2n - 2k}{2})
    \\ &\phantom{=asdfasdfasdf} \times {}_3F_{2} (-k,\frac{1-m}{2},-\frac{m}{2}; 1+n-k, 1- \frac{m+\l_3}{2} ; 1) \,.
\end{aligned} \label{eq:TFI_ex_infsum1}
\end{align}
The sum over $k$ for $N_{l,m,n}$ must be evaluated as limiting values; there are $\frac{0}{0}$ cancellations between ${n \choose k}$ and ${}_3 F_{2}$ for some $k > n$ which yield nonvanishing contributions to $N_{l,m,n}$.

The study of the function space becomes simpler when we take the generalised aligned limit $(v \cdot a) = 0$~\cite{Chen:2024bpf}, where the infinite sum in \eqref{eq:TFI_ex_infsum1} localises to $m=0$ and the coefficient $N$ simplifies;\footnote{The $N'_{l,n}$ coefficient could be rewritten as products of Gamma functions through reflection identities, which could be used to reorganise the integral in a closed form through Kamp\'{e} de F\'{e}riet Functions.}
\begin{align}
\begin{aligned}
    I [a,b] &= \frac{(4 \pi)^{D - \frac{1}{2}} \, i^{\l_3}}{2^{1 + 2(\l_1 + \l_2)} \Gamma(\l_1) \Gamma (\l_2) \Gamma (\l_3)} \frac{
    \sum_{l,n=0}^{\infty} \frac{N'_{l,n}}{l!\,n!} \left( 2i \frac{(a \cdot b)}{ b^2 } \right)^{l} \left( \frac{a^2}{b^2} \right)^{n}
    }{(v^2)^{\frac{1 + \l_3}{2}} (b^2)^{D - \frac{1}{2} - \l_1 - \l_2 - \frac{\l_3}{2}}} \,,
    \\ N'_{l,n} &:= \frac{\Gamma (D - \frac{1}{2} - \l_1 - \l_2 - \frac{\l_3}{2} + l + n) \Gamma ( \frac{\l_3}{2} ) \sin(\pi(D + l + 2n - \l_1 - \l_2 - \l_3))}{\Gamma (\frac{2 - D + 2 \l_2 + \l_3}{2}) \sin (\frac{\pi(D + 2n - 2 \l_2 - \l_3)}{2})}
    \\ &\phantom{=as} \times \Gamma (\frac{D - 2\l_1 - \l_3 + 2l + 2n}{2}) \Gamma (1 - D + \l_1 + \l_2 + \l_3 - l - n) \,.
\end{aligned} \label{eq:TFI_ex_infsum2}
\end{align}
The infinite sum \eqref{eq:TFI_ex_infsum2} can be simplified into a closed expression for some choices of parameters. We present some examples, where the superscript of $I$ denotes the choice of $(\l_1 , \l_2 , \l_3)$ in \eqref{eq:TFI_ex_infsum2}:
\begin{align}
    I^{(1,1,0)} &= \left\{ 
    \begin{aligned}
        & -\frac{2^{D-2} \pi^{D-1} \cos (\frac{\pi D}{2})}{\sqrt{v^2} (b^2)^{D - \frac{5}{2}}} \Gamma(\frac{3-D}{2}) \Gamma(\frac{D-2}{2}) && (a \cdot b) = 0  \\
        & \phantom{asdf} \times \Gamma(\frac{2D-5}{2}) \, {}_2 F_1 (\frac{2D - 5}{2}, \frac{D-2}{2}; D-2 ; \frac{a^2}{b^2}) \\
        & \frac{4 \pi^3}{\sqrt{v^2 b^2}} K(\frac{a^2}{b^2})&& (a \cdot b) = 0 \,,\, D = 3
    \end{aligned} \label{eq:TFI_2loop_ex1}
    \right. \\
    I^{(1,1,1)} &= \left\{ 
    \begin{aligned}
        & \frac{i 4^{D-3} \pi^{D+1} \Gamma(\frac{D-3}{2})}{v^2 \Gamma (\frac{5-D}{2}) \cos (\frac{\pi D}{2})} \frac{1}{[b^2 (b^2 - a^2)]^{\frac{D-3}{2}}} && (a \cdot b) = 0  \\
        & \frac{i 4^{D-3} \pi^{D+2} }{v^2 \Gamma (\frac{5-D}{2})^2 \cos^2 (\frac{\pi D}{2})} \frac{1}{(b^2 \mp i \sqrt{a^2 b^2})^{D-3}} && (a \cdot b) = \pm \sqrt{a^2 b^2}
    \end{aligned} \label{eq:TFI_2loop_ex2}
    \right.
\end{align}
The appearance of hypergeometric functions and elliptic integrals for some choices of parameters [\eqref{eq:TFI_2loop_ex1}] supports the conclusion of Picard\textendash Fuchs analysis that the underlying geometry for the integral family \eqref{eq:ibpset} cannot be the Riemann sphere.

One final remark is that while Bessel functions\textemdash a signature of exponential periods\textemdash appeared in the one-loop twisted Feynman integral \eqref{eq:TFI_ex_1loop}, the functions (hypergeometric functions ${}_2F_1$ and power-law behaviour $x^{-(D-3)}$) appearing in two-loop examples \eqref{eq:TFI_2loop_ex1} and \eqref{eq:TFI_2loop_ex2} are simpler in that the same functions also appear in usual Feynman integrals. 
It would be interesting to understand if the complexity of the function space of odd-loop twisted Feynman integrals is very different from that of even-loops,\footnote{Deformation to a twisted Feynman integral can be viewed as a Fourier transform. A momentum space expression returns to another momentum space expression after an even number of Fourier transforms.} or if the observed reduction of complexity in the function space is a coincidental behaviour specific to the considered example.

\section{Summary and Outlook} \label{sec:SUM}

Recently, there has been a growing interest in studying integrals similar to Feynman integrals that have an extra exponential factor in the integrand, mainly as a novel tool for tensor reduction~\cite{Feng:2022hyg} or for applications to post-Minkowskian dynamics of Kerr black holes~\cite{Kim:2024grz,Chen:2024bpf,Aoude:2025xxq}. It was shown that deforming the integrand by an exponential factor has a geometric interpretation; the loops in the corresponding Feynman diagram are ``twisted open'' and the virtual particle's spacetime trajectory corresponding to the loop is no longer a closed trajectory. Therefore, it is appropriate to call such integrals twisted Feynman integrals. A mathematical framework for describing twisted Feynman integrals which naturally incorporates this geometric viewpoint was constructed in sec.~\ref{sec:DEF}, and how conventional mathematical tools for evaluating Feynman integrals can and cannot handle their twisted cousin has been explored in sec.~\ref{sec:REP}.

A physical set-up for twisted Feynman integrals that has not been discussed is evaluating Fourier transforms of Feynman integrals, where the Fourier-transformed momentum is treated as an extra loop momentum~\cite{Brunello:2023fef}. 
A notable case where refined tools for evaluating twisted Feynman integrals can become useful is in the computation of gravitational waveforms in scattering encounters~\cite{Brunello:2024ibk,Brunello:2025cot,Brunello:2025eso} and the computation of dipole (quark-antiquark pair) scattering in QCD~\cite{Balitsky:1995ub}. 

For such phenomenological applications, obtaining the numerical value is often more important than understanding the functional properties of the integral. In this aspect, a natural future direction is to apply numerical evaluation techniques developed for standard Feynman integrals to their twisted counterparts. 
These techniques include various series-expansion methods, direct-integration approaches such as Monte Carlo integration, and the numerical integration of differential equations. 
A direct numerical integration based on the integrand parametrisation is expected to require additional care, since the exponential term breaks the homogeneity of the Symanzik polynomials and modifies the structure of their real-valued regions. 
In contrast, the numerical evaluation via differential equations will proceed in essentially the same way as for standard Feynman integrals. 

On the other hand, an efficient and stable numerical evaluation of an integral relies on understanding the analytic structures of the integrand, e.g., we do not expect the numerical integration to be reliable when the integral is expected to be divergent. 
This motivates the study of generalised Landau analysis; how can we determine the possible singular loci of twisted Feynman integrals? 
For example, the examples \eqref{eq:TFI_2loop_ex1} and \eqref{eq:TFI_2loop_ex2} for parameter choice $(a \cdot b) = 0$ develop singularities when approaching the point $b^2 = a^2$. Is it possible to know the existence of this singularity without the actual evaluation of the integrals?
Another topic that could benefit efficient numerical integration is the application of Mellin-Barnes parametrisation to twisted Feynman integrals.

One future direction is to establish a geometric perspective on twisted Feynman integrals from the language of gauge theory over discretized spaces~\cite{Hatcher,fradkin2015discretized,tong2018gt}. 
For instance, a realisation of a Feynman graph can be understood as a gauge representative, in which case the non-uniqueness of realising a Feynman graph describes the gauge ambiguities. 
In turn, there is a sense in which the twist $\a$ can be thought of as a Wilson loop or a holonomy due to a gauge connection over graphs.

\acknowledgments
The authors would like to thank 
Paul-Hermann Balduf, Francesco Calisto, Clifford Cheung, Stefano De Angelis, Lance Dixon, Vitaly Magerya, Sebastian Mizera, and Anna-Laura Sattelberger
for insightful discussions.
J.L wants to thank Christoph Dlapa, Hjalte Frellesvig, Johannes Henn, Pierpaolo Mastrolia, and Saiei-Jaeyeong Matsubara-Heo.
J.-H.K. is supported by the Department of Energy (Grant No.~DE-SC0011632) and by the Walter Burke Institute for Theoretical Physics.
J.L is supported by the Excellence Cluster ORIGINS funded by the Deutsche Forschungsgemeinschaft (DFG, German Research Foundation) under Germany’s Excellence Strategy – EXC-2094-390783311.
    
	\let\c\oldc
	\let\i\oldi
	\bibliography{references.bib}

\providecommand{\href}[2]{#2}\begingroup\raggedright\begin{thebibliography}{100}

\bibitem{Cvitanovic:1983eb}
P.~Cvitanovic, {\em {FIELD THEORY}}.
\newblock 1, 1983.

\bibitem{Veltman:1994wz}
M.~J.~G. Veltman, {\em {Diagrammatica: The Path to Feynman rules}}, vol.~4.
\newblock Cambridge University Press, 5, 2012.

\bibitem{Weinberg:1996kw}
S.~Weinberg, ``{What is quantum field theory, and what did we think it is?},''
  in {\em {Conference on Historical Examination and Philosophical Reflections
  on the Foundations of Quantum Field Theory}}, pp.~241--251.
\newblock 3, 1996.
\newblock \href{http://arxiv.org/abs/hep-th/9702027}{{\ttfamily
  arXiv:hep-th/9702027}}.

\bibitem{Passarino:1978jh}
G.~Passarino and M.~J.~G. Veltman, ``{One Loop Corrections for e+ e-
  Annihilation Into mu+ mu- in the Weinberg Model},''
  \href{http://dx.doi.org/10.1016/0550-3213(79)90234-7}{{\em Nucl. Phys. B}
  {\bfseries 160} (1979) 151--207}.

\bibitem{Chetyrkin:1981qh}
K.~G. Chetyrkin and F.~V. Tkachov, ``{Integration by parts: The algorithm to
  calculate $\beta$-functions in 4 loops},''
  \href{http://dx.doi.org/10.1016/0550-3213(81)90199-1}{{\em Nucl. Phys. B}
  {\bfseries 192} (1981) 159--204}.

\bibitem{Kotikov:1990kg}
A.~V. Kotikov, ``{Differential equations method: New technique for massive
  Feynman diagrams calculation},''
  \href{http://dx.doi.org/10.1016/0370-2693(91)90413-K}{{\em Phys. Lett. B}
  {\bfseries 254} (1991) 158--164}.

\bibitem{Bern:1994zx}
Z.~Bern, L.~J. Dixon, D.~C. Dunbar, and D.~A. Kosower, ``{One loop n point
  gauge theory amplitudes, unitarity and collinear limits},''
  \href{http://dx.doi.org/10.1016/0550-3213(94)90179-1}{{\em Nucl. Phys. B}
  {\bfseries 425} (1994) 217--260},
  \href{http://arxiv.org/abs/hep-ph/9403226}{{\ttfamily arXiv:hep-ph/9403226}}.

\bibitem{Bern:1994cg}
Z.~Bern, L.~J. Dixon, D.~C. Dunbar, and D.~A. Kosower, ``{Fusing gauge theory
  tree amplitudes into loop amplitudes},''
  \href{http://dx.doi.org/10.1016/0550-3213(94)00488-Z}{{\em Nucl. Phys. B}
  {\bfseries 435} (1995) 59--101},
  \href{http://arxiv.org/abs/hep-ph/9409265}{{\ttfamily arXiv:hep-ph/9409265}}.

\bibitem{Laporta:2000dsw}
S.~Laporta, ``{High-precision calculation of multiloop Feynman integrals by
  difference equations},''
  \href{http://dx.doi.org/10.1142/S0217751X00002159}{{\em Int. J. Mod. Phys. A}
  {\bfseries 15} (2000) 5087--5159},
  \href{http://arxiv.org/abs/hep-ph/0102033}{{\ttfamily arXiv:hep-ph/0102033}}.

\bibitem{Ossola:2006us}
G.~Ossola, C.~G. Papadopoulos, and R.~Pittau, ``{Reducing full one-loop
  amplitudes to scalar integrals at the integrand level},''
  \href{http://dx.doi.org/10.1016/j.nuclphysb.2006.11.012}{{\em Nucl. Phys. B}
  {\bfseries 763} (2007) 147--169},
  \href{http://arxiv.org/abs/hep-ph/0609007}{{\ttfamily arXiv:hep-ph/0609007}}.

\bibitem{Henn:2013pwa}
J.~M. Henn, ``{Multiloop integrals in dimensional regularization made
  simple},'' \href{http://dx.doi.org/10.1103/PhysRevLett.110.251601}{{\em Phys.
  Rev. Lett.} {\bfseries 110} (2013) 251601},
  \href{http://arxiv.org/abs/1304.1806}{{\ttfamily arXiv:1304.1806 [hep-th]}}.

\bibitem{Feng:2022hyg}
B.~Feng, ``{Generation function for one-loop tensor reduction},''
  \href{http://dx.doi.org/10.1088/1572-9494/aca253}{{\em Commun. Theor. Phys.}
  {\bfseries 75} no.~2, (2023) 025203},
  \href{http://arxiv.org/abs/2209.09517}{{\ttfamily arXiv:2209.09517
  [hep-ph]}}.

\bibitem{Chen:2024bpf}
G.~Chen, J.-W. Kim, and T.~Wang, ``{Systematic integral evaluation for
  spin-resummed binary dynamics},''
  \href{http://dx.doi.org/10.1103/PhysRevD.111.L021701}{{\em Phys. Rev. D}
  {\bfseries 111} no.~2, (2025) L021701},
  \href{http://arxiv.org/abs/2406.17658}{{\ttfamily arXiv:2406.17658
  [hep-th]}}.

\bibitem{Kim:2024grz}
J.-H. Kim, J.-W. Kim, and S.~Lee, ``{Massive twistor worldline in
  electromagnetic fields},''
  \href{http://dx.doi.org/10.1007/JHEP08(2024)080}{{\em JHEP} {\bfseries 08}
  (2024) 080}, \href{http://arxiv.org/abs/2405.17056}{{\ttfamily
  arXiv:2405.17056 [hep-th]}}.

\bibitem{Aoude:2025xxq}
R.~Aoude and A.~Helset, ``{Hidden simplicity in the scattering for neutron
  stars and black holes},'' \href{http://arxiv.org/abs/2509.04425}{{\ttfamily
  arXiv:2509.04425 [hep-th]}}.

\bibitem{GenSymGrav}
C.~Cheung, M.~Derda, J.-H. Kim, V.~Nevoa, I.~Rothstein, and N.~Shah,
  ``{Generalized symmetry in dynamical gravity},''
  \href{http://dx.doi.org/10.1007/JHEP10(2024)007}{{\em JHEP} {\bfseries 10}
  (2024) 007}, \href{http://arxiv.org/abs/2403.01837}{{\ttfamily
  arXiv:2403.01837 [hep-th]}}.

\bibitem{Wess:1974tw}
J.~Wess and B.~Zumino, ``{Supergauge Transformations in Four-Dimensions},''
  \href{http://dx.doi.org/10.1016/0550-3213(74)90355-1}{{\em Nucl. Phys. B}
  {\bfseries 70} (1974) 39--50}.

\bibitem{Salam:1974yz}
A.~Salam and J.~A. Strathdee, ``{Supergauge Transformations},''
  \href{http://dx.doi.org/10.1016/0550-3213(74)90537-9}{{\em Nucl. Phys. B}
  {\bfseries 76} (1974) 477--482}.

\bibitem{Rattazzi:2008pe}
R.~Rattazzi, V.~S. Rychkov, E.~Tonni, and A.~Vichi, ``{Bounding scalar operator
  dimensions in 4D CFT},''
  \href{http://dx.doi.org/10.1088/1126-6708/2008/12/031}{{\em JHEP} {\bfseries
  12} (2008) 031}, \href{http://arxiv.org/abs/0807.0004}{{\ttfamily
  arXiv:0807.0004 [hep-th]}}.

\bibitem{Arkani-Hamed:2018ign}
N.~Arkani-Hamed, Y.-T. Huang, and S.-H. Shao, ``{On the Positive Geometry of
  Conformal Field Theory},''
  \href{http://dx.doi.org/10.1007/JHEP06(2019)124}{{\em JHEP} {\bfseries 06}
  (2019) 124}, \href{http://arxiv.org/abs/1812.07739}{{\ttfamily
  arXiv:1812.07739 [hep-th]}}.

\bibitem{Mastrolia:2018uzb}
P.~Mastrolia and S.~Mizera, ``{Feynman Integrals and Intersection Theory},''
  \href{http://dx.doi.org/10.1007/JHEP02(2019)139}{{\em JHEP} {\bfseries 02}
  (2019) 139}, \href{http://arxiv.org/abs/1810.03818}{{\ttfamily
  arXiv:1810.03818 [hep-th]}}.

\bibitem{Marcolli:2009zy}
M.~Marcolli, ``{Feynman integrals and motives},''
  \href{http://arxiv.org/abs/0907.0321}{{\ttfamily arXiv:0907.0321 [math-ph]}}.

\bibitem{Baikov:1996iu}
P.~A. Baikov, ``{Explicit solutions of the multiloop integral recurrence
  relations and its application},''
  \href{http://dx.doi.org/10.1016/S0168-9002(97)00126-5}{{\em Nucl. Instrum.
  Meth. A} {\bfseries 389} (1997) 347--349},
  \href{http://arxiv.org/abs/hep-ph/9611449}{{\ttfamily arXiv:hep-ph/9611449}}.

\bibitem{Newman:1965my-kerrmetric}
E.~T. Newman, R.~Couch, K.~Chinnapared, A.~Exton, A.~Prakash, and R.~Torrence,
  ``{Metric of a rotating, charged mass},''
  \href{http://dx.doi.org/10.1063/1.1704351}{{\em J. Math. Phys.} {\bfseries 6}
  (1965) 918--919}.

\bibitem{talbot1969newman}
C.~Talbot, ``Newman-penrose approach to twisting degenerate metrics,'' {\em
  Communications in Mathematical Physics} {\bfseries 13} no.~1, (1969) 45--61.

\bibitem{Drake:1998gf}
S.~P. Drake and P.~Szekeres, ``{Uniqueness of the Newman–Janis Algorithm in
  Generating the Kerr–Newman Metric},''
  \href{http://dx.doi.org/10.1023/A:1001920232180}{{\em Gen. Rel. Grav.}
  {\bfseries 32} (2000) 445--458},
  \href{http://arxiv.org/abs/gr-qc/9807001}{{\ttfamily arXiv:gr-qc/9807001}}.

\bibitem{Gurses:1975vu}
{G\"urses, Metin and G\"ursey, Feza}, ``{Lorentz Covariant Treatment of the
  {Kerr-Schild} Metric},'' \href{http://dx.doi.org/10.1063/1.522480}{{\em J.
  Math. Phys.} {\bfseries 16} (1975) 2385}.

\bibitem{flaherty1976hermitian}
E.~J. Flaherty, {\em Hermitian and K{\"a}hlerian geometry in relativity}.
\newblock Springer, 1976.

\bibitem{grg207flaherty}
E.~J. Flaherty~Jr., ``Complex variables in relativity,'' in {\em General
  Relativity and Gravitation: One Hundred Years After the Birth of Albert
  Einstein}, vol.~2, pp.~207--239, International Society on General Relativity
  and Gravitation.
\newblock 1980.

\bibitem{Rajan:2016zmq}
D.~Rajan and M.~Visser, ``{Cartesian {Kerr-Schild} variation on the
  {{Newman-Janis}} trick},''
  \href{http://dx.doi.org/10.1142/S021827181750167X}{{\em Int. J. Mod. Phys. D}
  {\bfseries 26} no.~14, (2017) 1750167},
  \href{http://arxiv.org/abs/1601.03532}{{\ttfamily arXiv:1601.03532 [gr-qc]}}.

\bibitem{giampieri1990introducing}
G.~Giampieri, ``Introducing angular momentum into a black hole using complex
  variables,'' {\em Gravity Research Foundation} (1990) .

\bibitem{Newman:1965tw-janis}
E.~T. Newman and A.~I. Janis, ``{Note on the {Kerr} spinning particle
  metric},'' \href{http://dx.doi.org/10.1063/1.1704350}{{\em J. Math. Phys.}
  {\bfseries 6} (1965) 915--917}.

\bibitem{penrose1967twistoralgebra}
R.~Penrose, ``Twistor algebra,'' {\em Journal of Mathematical Physics}
  {\bfseries 8} no.~2, (1967) 345--366.

\bibitem{newman1988remarkable}
E.~T. Newman, ``The remarkable efficacy of complex methods in general
  relativity,'' {\em Highlights in Gravitation and Cosmology} {\bfseries 67}
  (1988) .

\bibitem{Newman:1973afx}
E.~T. Newman, ``{Complex coordinate transformations and the Schwarzschild-Kerr
  metrics},'' \href{http://dx.doi.org/10.1063/1.1666393}{{\em J. Math. Phys.}
  {\bfseries 14} no.~6, (1973) 774}.

\bibitem{Newman:2002mk}
E.~T. Newman, ``{On a classical, geometric origin of magnetic moments, spin
  angular momentum and the {Dirac} gyromagnetic ratio},''
  \href{http://dx.doi.org/10.1103/PhysRevD.65.104005}{{\em Phys. Rev. D}
  {\bfseries 65} (2002) 104005},
  \href{http://arxiv.org/abs/gr-qc/0201055}{{\ttfamily arXiv:gr-qc/0201055}}.

\bibitem{Newman:1973yu}
E.~T. Newman, ``{{Maxwell}'s equations and complex {Minkowski} space},''
  \href{http://dx.doi.org/10.1063/1.1666160}{{\em J. Math. Phys.} {\bfseries
  14} (1973) 102--103}.

\bibitem{newman1974curiosity}
E.~T. Newman and J.~Winicour, ``{A curiosity concerning angular momentum},''
  \href{http://dx.doi.org/10.1063/1.1666761}{{\em J. Math. Phys.} {\bfseries
  15} (1974) 1113--1115}.

\bibitem{newman1974collection}
{Newman, Ezra T}, ``{Complex space-time {\&} some curious consequences},'' in
  {\em {Feldafing Conference of the Max-Planck Inst. on Quantum Theory and the
  Structure of Space-time}}, pp.~117--127.
\newblock 1974.

\bibitem{newman1973complex}
E.~T. Newman, ``Complex coordinate transformations and the
  {Schwarzschild}-{Kerr} metrics,'' {\em Journal of Mathematical Physics}
  {\bfseries 14} no.~6, (1973) 774--776.

\bibitem{newman2004maxwell}
E.~T. Newman, ``Maxwell fields and shear-free null geodesic congruences,'' {\em
  Classical and Quantum Gravity} {\bfseries 21} no.~13, (2004) 3197.

\bibitem{newman1976heaven}
E.~T. Newman, ``Heaven and its properties,'' {\em General Relativity and
  Gravitation} {\bfseries 7} no.~1, (1976) 107--111.

\bibitem{ko1981theory}
M.~Ko, M.~Ludvigsen, E.~Newman, and K.~Tod, ``The theory of
  $\mathscr{H}$-space,'' {\em Physics Reports} {\bfseries 71} no.~2, (1981)
  51--139.

\bibitem{sst-asym}
J.-H. Kim, ``{Asymptotic spinspacetime},''
  \href{http://dx.doi.org/10.1103/PhysRevD.111.105011}{{\em Phys. Rev. D}
  {\bfseries 111} no.~10, (2025) 105011},
  \href{http://arxiv.org/abs/2309.11886}{{\ttfamily arXiv:2309.11886
  [hep-th]}}.

\bibitem{Arkani-Hamed:2017jhn}
N.~Arkani-Hamed, T.-C. Huang, and Y.-t. Huang, ``{Scattering amplitudes for all
  masses and spins},'' \href{http://dx.doi.org/10.1007/JHEP11(2021)070}{{\em
  JHEP} {\bfseries 11} (2021) 070},
  \href{http://arxiv.org/abs/1709.04891}{{\ttfamily arXiv:1709.04891
  [hep-th]}}.

\bibitem{Guevara:2018wpp}
A.~Guevara, A.~Ochirov, and J.~Vines, ``{Scattering of Spinning Black Holes
  from Exponentiated Soft Factors},''
  \href{http://dx.doi.org/10.1007/JHEP09(2019)056}{{\em JHEP} {\bfseries 09}
  (2019) 056}, \href{http://arxiv.org/abs/1812.06895}{{\ttfamily
  arXiv:1812.06895 [hep-th]}}.

\bibitem{Guevara:2019fsj}
A.~Guevara, A.~Ochirov, and J.~Vines, ``{Black-hole scattering with general
  spin directions from minimal-coupling amplitudes},''
  \href{http://dx.doi.org/10.1103/PhysRevD.100.104024}{{\em Phys. Rev. D}
  {\bfseries 100} no.~10, (2019) 104024},
  \href{http://arxiv.org/abs/1906.10071}{{\ttfamily arXiv:1906.10071
  [hep-th]}}.

\bibitem{chkl2019}
M.-Z. Chung, Y.-T. Huang, J.-W. Kim, and S.~Lee, ``{The simplest massive
  S-matrix: from minimal coupling to Black Holes},''
  \href{http://dx.doi.org/10.1007/JHEP04(2019)156}{{\em JHEP} {\bfseries 04}
  (2019) 156}, \href{http://arxiv.org/abs/1812.08752}{{\ttfamily
  arXiv:1812.08752 [hep-th]}}.

\bibitem{aho2020}
N.~Arkani-Hamed, Y.-t. Huang, and D.~O'Connell, ``{Kerr black holes as
  elementary particles},''
  \href{http://dx.doi.org/10.1007/JHEP01(2020)046}{{\em JHEP} {\bfseries 01}
  (2020) 046}, \href{http://arxiv.org/abs/1906.10100}{{\ttfamily
  arXiv:1906.10100 [hep-th]}}.

\bibitem{Adamo:2014baa}
T.~Adamo and E.~T. Newman, ``{The Kerr-Newman metric: A Review},''
  \href{http://dx.doi.org/10.4249/scholarpedia.31791}{{\em Scholarpedia}
  {\bfseries 9} (2014) 31791}, \href{http://arxiv.org/abs/1410.6626}{{\ttfamily
  arXiv:1410.6626 [gr-qc]}}.

\bibitem{Erbin:2016lzq}
H.~Erbin, ``{ {Janis-Newman} algorithm: generating rotating and NUT charged
  black holes},'' \href{http://dx.doi.org/10.3390/universe3010019}{{\em
  Universe} {\bfseries 3} no.~1, (2017) 19},
  \href{http://arxiv.org/abs/1701.00037}{{\ttfamily arXiv:1701.00037 [gr-qc]}}.

\bibitem{nja}
J.-H. Kim, ``{Newman-Janis Algorithm from Taub-NUT Instantons},''
  \href{http://arxiv.org/abs/2412.19611}{{\ttfamily arXiv:2412.19611 [gr-qc]}}.

\bibitem{Dhani:2024jja}
A.~Dhani, S.~H. V{\"o}lkel, A.~Buonanno, H.~Estelles, J.~Gair, H.~P. Pfeiffer,
  L.~Pompili, and A.~Toubiana, ``{Systematic Biases in Estimating the
  Properties of Black Holes Due to Inaccurate Gravitational-Wave Models},''
  \href{http://dx.doi.org/10.1103/5pks-qz6b}{{\em Phys. Rev. X} {\bfseries 15}
  no.~3, (2025) 031036}, \href{http://arxiv.org/abs/2404.05811}{{\ttfamily
  arXiv:2404.05811 [gr-qc]}}.

\bibitem{LIGOScientific:2025rsn}
{\bfseries LIGO Scientific, VIRGO, KAGRA} {\bfseries Collaboration}, A.~G. Abac
  { et~al.}, ``{GW231123: a Binary Black Hole Merger with Total Mass 190-265
  $M_{\odot}$},'' \href{http://arxiv.org/abs/2507.08219}{{\ttfamily
  arXiv:2507.08219 [astro-ph.HE]}}.

\bibitem{Aoude:2022thd}
R.~Aoude, K.~Haddad, and A.~Helset, ``{Classical Gravitational
  Spinning-Spinless Scattering at O(G2S{\ensuremath{\infty}})},''
  \href{http://dx.doi.org/10.1103/PhysRevLett.129.141102}{{\em Phys. Rev.
  Lett.} {\bfseries 129} no.~14, (2022) 141102},
  \href{http://arxiv.org/abs/2205.02809}{{\ttfamily arXiv:2205.02809
  [hep-th]}}.

\bibitem{Damgaard:2022jem}
P.~H. Damgaard, J.~Hoogeveen, A.~Luna, and J.~Vines, ``{Scattering angles in
  Kerr metrics},'' \href{http://dx.doi.org/10.1103/PhysRevD.106.124030}{{\em
  Phys. Rev. D} {\bfseries 106} no.~12, (2022) 124030},
  \href{http://arxiv.org/abs/2208.11028}{{\ttfamily arXiv:2208.11028
  [hep-th]}}.

\bibitem{Bohnenblust:2024hkw}
L.~Bohnenblust, L.~Cangemi, H.~Johansson, and P.~Pichini, ``{Binary Kerr
  black-hole scattering at 2PM from quantum higher-spin Compton},''
  \href{http://dx.doi.org/10.1007/JHEP07(2025)261}{{\em JHEP} {\bfseries 07}
  (2025) 261}, \href{http://arxiv.org/abs/2410.23271}{{\ttfamily
  arXiv:2410.23271 [hep-th]}}.

\bibitem{Ramos-Buades:2023ehm}
A.~Ramos-Buades, A.~Buonanno, H.~Estell{\'e}s, M.~Khalil, D.~P. Mihaylov,
  S.~Ossokine, L.~Pompili, and M.~Shiferaw, ``{Next generation of accurate and
  efficient multipolar precessing-spin effective-one-body waveforms for binary
  black holes},'' \href{http://dx.doi.org/10.1103/PhysRevD.108.124037}{{\em
  Phys. Rev. D} {\bfseries 108} no.~12, (2023) 124037},
  \href{http://arxiv.org/abs/2303.18046}{{\ttfamily arXiv:2303.18046 [gr-qc]}}.

\bibitem{Khalil:2023kep}
M.~Khalil, A.~Buonanno, H.~Estelles, D.~P. Mihaylov, S.~Ossokine, L.~Pompili,
  and A.~Ramos-Buades, ``{Theoretical groundwork supporting the precessing-spin
  two-body dynamics of the effective-one-body waveform models SEOBNRv5},''
  \href{http://dx.doi.org/10.1103/PhysRevD.108.124036}{{\em Phys. Rev. D}
  {\bfseries 108} no.~12, (2023) 124036},
  \href{http://arxiv.org/abs/2303.18143}{{\ttfamily arXiv:2303.18143 [gr-qc]}}.

\bibitem{Buonanno:2024vkx}
A.~Buonanno, G.~U. Jakobsen, and G.~Mogull, ``{Post-Minkowskian theory meets
  the spinning effective-one-body approach for two-body scattering},''
  \href{http://dx.doi.org/10.1103/PhysRevD.110.044038}{{\em Phys. Rev. D}
  {\bfseries 110} no.~4, (2024) 044038},
  \href{http://arxiv.org/abs/2402.12342}{{\ttfamily arXiv:2402.12342 [gr-qc]}}.

\bibitem{Chung:2018kqs}
M.-Z. Chung, Y.-T. Huang, J.-W. Kim, and S.~Lee, ``{The simplest massive
  S-matrix: from minimal coupling to Black Holes},''
  \href{http://dx.doi.org/10.1007/JHEP04(2019)156}{{\em JHEP} {\bfseries 04}
  (2019) 156}, \href{http://arxiv.org/abs/1812.08752}{{\ttfamily
  arXiv:1812.08752 [hep-th]}}.

\bibitem{Chung:2019yfs}
M.-Z. Chung, Y.-T. Huang, and J.-W. Kim, ``{Kerr-Newman stress-tensor from
  minimal coupling},'' \href{http://dx.doi.org/10.1007/JHEP12(2020)103}{{\em
  JHEP} {\bfseries 12} (2020) 103},
  \href{http://arxiv.org/abs/1911.12775}{{\ttfamily arXiv:1911.12775
  [hep-th]}}.

\bibitem{Arkani-Hamed:2019ymq}
N.~Arkani-Hamed, Y.-t. Huang, and D.~O'Connell, ``{Kerr black holes as
  elementary particles},''
  \href{http://dx.doi.org/10.1007/JHEP01(2020)046}{{\em JHEP} {\bfseries 01}
  (2020) 046}, \href{http://arxiv.org/abs/1906.10100}{{\ttfamily
  arXiv:1906.10100 [hep-th]}}.

\bibitem{Chiodaroli:2021eug}
M.~Chiodaroli, H.~Johansson, and P.~Pichini, ``{Compton black-hole scattering
  for s {\ensuremath{\leq}} 5/2},''
  \href{http://dx.doi.org/10.1007/JHEP02(2022)156}{{\em JHEP} {\bfseries 02}
  (2022) 156}, \href{http://arxiv.org/abs/2107.14779}{{\ttfamily
  arXiv:2107.14779 [hep-th]}}.

\bibitem{Cangemi:2022bew}
L.~Cangemi, M.~Chiodaroli, H.~Johansson, A.~Ochirov, P.~Pichini, and
  E.~Skvortsov, ``{Kerr Black Holes From Massive Higher-Spin Gauge Symmetry},''
  \href{http://dx.doi.org/10.1103/PhysRevLett.131.221401}{{\em Phys. Rev.
  Lett.} {\bfseries 131} no.~22, (2023) 221401},
  \href{http://arxiv.org/abs/2212.06120}{{\ttfamily arXiv:2212.06120
  [hep-th]}}.

\bibitem{Bjerrum-Bohr:2023jau}
N.~E.~J. Bjerrum-Bohr, G.~Chen, and M.~Skowronek, ``{Classical spin
  gravitational Compton scattering},''
  \href{http://dx.doi.org/10.1007/JHEP06(2023)170}{{\em JHEP} {\bfseries 06}
  (2023) 170}, \href{http://arxiv.org/abs/2302.00498}{{\ttfamily
  arXiv:2302.00498 [hep-th]}}.

\bibitem{Bjerrum-Bohr:2023iey}
N.~E.~J. Bjerrum-Bohr, G.~Chen, and M.~Skowronek, ``{Covariant Compton
  Amplitudes in Gravity with Classical Spin},''
  \href{http://dx.doi.org/10.1103/PhysRevLett.132.191603}{{\em Phys. Rev.
  Lett.} {\bfseries 132} no.~19, (2024) 191603},
  \href{http://arxiv.org/abs/2309.11249}{{\ttfamily arXiv:2309.11249
  [hep-th]}}.

\bibitem{Guevara:2020xjx}
A.~Guevara, B.~Maybee, A.~Ochirov, D.~O'connell, and J.~Vines, ``{A worldsheet
  for Kerr},'' \href{http://dx.doi.org/10.1007/JHEP03(2021)201}{{\em JHEP}
  {\bfseries 03} (2021) 201}, \href{http://arxiv.org/abs/2012.11570}{{\ttfamily
  arXiv:2012.11570 [hep-th]}}.

\bibitem{Kim:2023aff}
J.-H. Kim and S.~Lee, ``{Symplectic Perturbation Theory in Massive Ambitwistor
  Space: A Zig-Zag Theory of Massive Spinning Particles},''
  \href{http://arxiv.org/abs/2301.06203}{{\ttfamily arXiv:2301.06203
  [hep-th]}}.

\bibitem{Lynden-Bell:2002dvr}
D.~Lynden-Bell, ``{A magic electromagnetic field},'' in {\em {Stellar
  Astrophysical Fluid Dynamics}}, ch.~25, pp.~369--376.
\newblock Cambridge University Press, 2003.
\newblock {arXiv:astro-ph/0207064}.

\bibitem{Kim:2021rda}
J.-H. Kim, J.-W. Kim, and S.~Lee, ``{The relativistic spherical top as a
  massive twistor},'' \href{http://dx.doi.org/10.1088/1751-8121/ac11be}{{\em J.
  Phys. A} {\bfseries 54} no.~33, (2021) 335203},
  \href{http://arxiv.org/abs/2102.07063}{{\ttfamily arXiv:2102.07063
  [hep-th]}}.

\bibitem{probe-nj}
J.-H. Kim, ``Note on the kerr spinning-particle equations of motion.'' To
  Appear.

\bibitem{sodual}
J.-H. Kim, ``Kerr effective action from orbit-spin duality.'' To Appear.

\bibitem{gde}
J.-H. Kim, ``{Geodesic deviation to all orders via a tangent bundle
  formalism},'' \href{http://arxiv.org/abs/2509.23600}{{\ttfamily
  arXiv:2509.23600 [gr-qc]}}.

\bibitem{hawking1977gravitational}
S.~W. Hawking, ``{Gravitational Instantons},''
  \href{http://dx.doi.org/10.1016/0375-9601(77)90386-3}{{\em Phys. Lett. A}
  {\bfseries 60} (1977) 81}.

\bibitem{Gibbons:1978tef}
G.~W. Gibbons and S.~W. Hawking, ``{Gravitational Multi-Instantons},''
  \href{http://dx.doi.org/10.1016/0370-2693(78)90478-1}{{\em Phys. Lett. B}
  {\bfseries 78} (1978) 430}.

\bibitem{Taub:1950ez}
A.~H. Taub, ``{Empty space-times admitting a three parameter group of
  motions},'' \href{http://dx.doi.org/10.2307/1969567}{{\em Annals Math.}
  {\bfseries 53} (1951) 472--490}.

\bibitem{Newman:1963yy}
E.~Newman, L.~Tamburino, and T.~Unti, ``{Empty space generalization of the
  Schwarzschild metric},'' \href{http://dx.doi.org/10.1063/1.1704018}{{\em J.
  Math. Phys.} {\bfseries 4} (1963) 915}.

\bibitem{Misner:1963flatter}
C.~W. Misner, ``{The Flatter regions of Newman, Unti and Tamburino's
  generalized Schwarzschild space},''
  \href{http://dx.doi.org/10.1063/1.1704019}{{\em J. Math. Phys.} {\bfseries 4}
  (1963) 924--938}.

\bibitem{Bonnor:1969ala}
W.~B. Bonnor, ``A new interpretation of the {{NUT}} metric in general
  relativity,'' \href{http://dx.doi.org/10.1017/s0305004100044807}{{\em Math.
  Proc. Cambridge Phil. Soc.} {\bfseries 66} no.~1, (1969) 145--151}.

\bibitem{sackfield1971physical}
A.~Sackfield, ``Physical interpretation of {NUT} metric,'' in {\em Mathematical
  Proceedings of the Cambridge Philosophical Society}, vol.~1, pp.~89--94,
  Cambridge University Press.
\newblock 1971.

\bibitem{demianski1966combined}
M.~Demianski and E.~T. Newman, ``{A combined Kerr-NUT solution of the Einstein
  field equations},'' {\em Bulletin de l'Academie Polonaise des Sciences, Serie
  des Sciences Mathematiques, Astronomiques et Physiques} {\bfseries 14} (12,
  1966) 653--657.

\bibitem{Plebanski:1975xfb}
J.~F. Pleba{\'n}ski, ``{A class of solutions of Einstein-Maxwell equations},''
  \href{http://dx.doi.org/10.1016/0003-4916(75)90145-1}{{\em Annals Phys.}
  {\bfseries 90} no.~1, (1975) 196--255}.

\bibitem{dowker1974nut}
J.~Dowker, ``The {NUT} solution as a gravitational dyon,'' {\em General
  Relativity and Gravitation} {\bfseries 5} (1974) 603--613.

\bibitem{Griffiths:2009dfa}
J.~B. Griffiths and J.~Podolsky,
  \href{http://dx.doi.org/10.1017/CBO9780511635397}{{\em {Exact Space-Times in
  Einstein's General Relativity}}}.
\newblock Cambridge Monographs on Mathematical Physics. Cambridge University
  Press, Cambridge, 2009.

\bibitem{monteiro2014black}
R.~Monteiro, D.~O’Connell, and C.~D. White, ``Black holes and the double
  copy,'' {\em Journal of High Energy Physics} {\bfseries 2014} no.~12, (2014)
  1--23.

\bibitem{Schubert:2001he}
C.~Schubert, ``{Perturbative quantum field theory in the string inspired
  formalism},'' \href{http://dx.doi.org/10.1016/S0370-1573(01)00013-8}{{\em
  Phys. Rept.} {\bfseries 355} (2001) 73--234},
  \href{http://arxiv.org/abs/hep-th/0101036}{{\ttfamily arXiv:hep-th/0101036}}.

\bibitem{Collins:1984xc}
J.~C. Collins, \href{http://dx.doi.org/10.1017/9781009401807}{{\em
  {Renormalization : An Introduction to Renormalization, the Renormalization
  Group and the Operator-Product Expansion}}}, vol.~26 of {\em Cambridge
  Monographs on Mathematical Physics}.
\newblock Cambridge University Press, Cambridge, 1984.

\bibitem{Hatcher}
A.~Hatcher, {\em {Algebraic topology}}.
\newblock Cambridge Univ. Press, Cambridge, 2000.
\newblock \url{https://cds.cern.ch/record/478079}.

\bibitem{griffiths2023introduction}
D.~J. Griffiths, {\em Introduction to electrodynamics}.
\newblock Cambridge University Press, 2023.

\bibitem{romer1982voltmeters}
R.~H. Romer, ``What do voltmeters measure?: Faraday’s law in a multiply
  connected region,'' {\em Am. J. Phys} {\bfseries 50} no.~12, (1982) 1089--93.

\bibitem{Kontsevich2001}
M.~Kontsevich and D.~Zagier, {\em Periods},
  \href{http://dx.doi.org/10.1007/978-3-642-56478-9_39}{pp.~771--808}.
\newblock Springer Berlin Heidelberg, Berlin, Heidelberg, 2001.
\newblock \url{https://doi.org/10.1007/978-3-642-56478-9_39}.

\bibitem{10.1155/S107379280313142X}
P.~Belkale and P.~Brosnan, ``Periods and igusa local zeta functions,''
  \href{http://dx.doi.org/10.1155/S107379280313142X}{{\em International
  Mathematics Research Notices} {\bfseries 2003} no.~49, (01, 2003)
  2655--2670}.

\bibitem{Bogner:2007mn}
C.~Bogner and S.~Weinzierl, ``{Periods and Feynman integrals},''
  \href{http://dx.doi.org/10.1063/1.3106041}{{\em J. Math. Phys.} {\bfseries
  50} (2009) 042302}, \href{http://arxiv.org/abs/0711.4863}{{\ttfamily
  arXiv:0711.4863 [hep-th]}}.

\bibitem{brown2010periodsfeynmanintegrals}
F.~C.~S. Brown, ``On the periods of some feynman integrals,''
  \href{http://arxiv.org/abs/0910.0114}{{\ttfamily arXiv:0910.0114 [math.AG]}}.
  \url{https://arxiv.org/abs/0910.0114}.

\bibitem{dlmf}
{NIST Digital Library of Mathematical Functions}. \url{https://dlmf.nist.gov/}.

\bibitem{Brunello:2023fef}
G.~Brunello, G.~Crisanti, M.~Giroux, P.~Mastrolia, and S.~Smith, ``{Fourier
  calculus from intersection theory},''
  \href{http://dx.doi.org/10.1103/PhysRevD.109.094047}{{\em Phys. Rev. D}
  {\bfseries 109} no.~9, (2024) 094047},
  \href{http://arxiv.org/abs/2311.14432}{{\ttfamily arXiv:2311.14432
  [hep-th]}}.

\bibitem{Landau:1959fi}
L.~D. Landau, ``{On the Analytic Properties of Vertex Parts in Quantum Field
  Theory},'' \href{http://dx.doi.org/10.1016/B978-0-08-010586-4.50103-6}{{\em
  Zh. Eksp. Teor. Fiz.} {\bfseries 37} no.~1, (1960) 62--70}.

\bibitem{Flieger:2022xyq}
W.~Flieger and W.~J. Torres~Bobadilla, ``{Landau and leading singularities in
  arbitrary space-time dimensions},''
  \href{http://dx.doi.org/10.1140/epjp/s13360-024-05796-7}{{\em Eur. Phys. J.
  Plus} {\bfseries 139} no.~11, (2024) 1022},
  \href{http://arxiv.org/abs/2210.09872}{{\ttfamily arXiv:2210.09872
  [hep-th]}}.

\bibitem{Brammer:2025rqo}
D.~Brammer, H.~Frellesvig, R.~Morales, and M.~Wilhelm, ``{Classification of
  Feynman integral geometries for black-hole scattering at 5PM order},''
  \href{http://arxiv.org/abs/2505.10274}{{\ttfamily arXiv:2505.10274
  [hep-th]}}.

\bibitem{Lee:2013hzt}
R.~N. Lee and A.~A. Pomeransky, ``{Critical points and number of master
  integrals},'' \href{http://dx.doi.org/10.1007/JHEP11(2013)165}{{\em JHEP}
  {\bfseries 11} (2013) 165}, \href{http://arxiv.org/abs/1308.6676}{{\ttfamily
  arXiv:1308.6676 [hep-ph]}}.

\bibitem{Kotikov:1991hm}
A.~V. Kotikov, ``{Differential equations method: The Calculation of vertex type
  Feynman diagrams},''
  \href{http://dx.doi.org/10.1016/0370-2693(91)90834-D}{{\em Phys. Lett. B}
  {\bfseries 259} (1991) 314--322}.

\bibitem{Kotikov:1991pm}
A.~V. Kotikov, ``{Differential equation method: The Calculation of N point
  Feynman diagrams},''
  \href{http://dx.doi.org/10.1016/0370-2693(91)90536-Y}{{\em Phys. Lett. B}
  {\bfseries 267} (1991) 123--127}. [Erratum: Phys.Lett.B 295, 409--409
  (1992)].

\bibitem{Gehrmann:1999as}
T.~Gehrmann and E.~Remiddi, ``{Differential equations for two-loop four-point
  functions},'' \href{http://dx.doi.org/10.1016/S0550-3213(00)00223-6}{{\em
  Nucl. Phys. B} {\bfseries 580} (2000) 485--518},
  \href{http://arxiv.org/abs/hep-ph/9912329}{{\ttfamily arXiv:hep-ph/9912329}}.

\bibitem{Brunello:2024ibk}
G.~Brunello and S.~De~Angelis, ``{An improved framework for computing
  waveforms},'' \href{http://dx.doi.org/10.1007/JHEP07(2024)062}{{\em JHEP}
  {\bfseries 07} (2024) 062}, \href{http://arxiv.org/abs/2403.08009}{{\ttfamily
  arXiv:2403.08009 [hep-th]}}.

\bibitem{Hu:2024kch}
C.~Hu, B.~Feng, J.~Shen, and Y.~Zhang, ``{General one-loop generating function
  by integration-by-part relations},''
  \href{http://dx.doi.org/10.1103/PhysRevD.111.076026}{{\em Phys. Rev. D}
  {\bfseries 111} no.~7, (2025) 076026},
  \href{http://arxiv.org/abs/2403.16040}{{\ttfamily arXiv:2403.16040
  [hep-ph]}}.

\bibitem{Chen:2025gqu}
X.~Chen, B.~Feng, and L.~Zhang, ``{Tensor Reduction of Sunset by Generating
  Function},'' \href{http://arxiv.org/abs/2509.18730}{{\ttfamily
  arXiv:2509.18730 [hep-th]}}.

\bibitem{Feng:2025leo}
B.~Feng, X.~Li, Y.~Liu, Y.-Q. Ma, and Y.~Zhang, ``{Symbolic Reduction of
  Multi-loop Feynman Integrals via Generating Functions},''
  \href{http://arxiv.org/abs/2509.21769}{{\ttfamily arXiv:2509.21769
  [hep-ph]}}.

\bibitem{Smirnov:2010hn}
A.~V. Smirnov and A.~V. Petukhov, ``{The Number of Master Integrals is
  Finite},'' \href{http://dx.doi.org/10.1007/s11005-010-0450-0}{{\em Lett.
  Math. Phys.} {\bfseries 97} (2011) 37--44},
  \href{http://arxiv.org/abs/1004.4199}{{\ttfamily arXiv:1004.4199 [hep-th]}}.

\bibitem{Bitoun:2017nre}
T.~Bitoun, C.~Bogner, R.~P. Klausen, and E.~Panzer, ``{Feynman integral
  relations from parametric annihilators},''
  \href{http://dx.doi.org/10.1007/s11005-018-1114-8}{{\em Lett. Math. Phys.}
  {\bfseries 109} no.~3, (2019) 497--564},
  \href{http://arxiv.org/abs/1712.09215}{{\ttfamily arXiv:1712.09215
  [hep-th]}}.

\bibitem{goncharov2013simpleconstructiongrassmannianpolylogarithms}
A.~B. Goncharov, ``A simple construction of grassmannian polylogarithms,''
  2013.
\newblock \url{https://arxiv.org/abs/0908.2238}.

\bibitem{Peraro:2019svx}
T.~Peraro, ``{$\text{FiniteFlow}$: multivariate functional reconstruction using
  finite fields and dataflow graphs},''
  \href{http://dx.doi.org/10.1007/JHEP07(2019)031}{{\em JHEP} {\bfseries 07}
  (2019) 031}, \href{http://arxiv.org/abs/1905.08019}{{\ttfamily
  arXiv:1905.08019 [hep-ph]}}.

\bibitem{Brunello:2025cot}
G.~Brunello, V.~Chestnov, G.~Crisanti, M.~Giroux, and S.~Smith,
  ``{Gravitational waveforms from restriction theory and rapid-decay
  homology},'' \href{http://arxiv.org/abs/2510.26874}{{\ttfamily
  arXiv:2510.26874 [hep-th]}}.

\bibitem{Brunello:2025eso}
G.~Brunello, S.~De~Angelis, and D.~A. Kosower, ``{Analytic One-loop Scattering
  Waveform in General Relativity},''
  \href{http://arxiv.org/abs/2511.05412}{{\ttfamily arXiv:2511.05412
  [hep-th]}}.

\bibitem{Balitsky:1995ub}
I.~Balitsky, ``{Operator expansion for high-energy scattering},''
  \href{http://dx.doi.org/10.1016/0550-3213(95)00638-9}{{\em Nucl. Phys. B}
  {\bfseries 463} (1996) 99--160},
  \href{http://arxiv.org/abs/hep-ph/9509348}{{\ttfamily arXiv:hep-ph/9509348}}.

\bibitem{fradkin2015discretized}
K.~Sun, K.~Kumar, and E.~Fradkin, ``Discretized abelian chern-simons gauge
  theory on arbitrary graphs,'' {\em Physical Review B} {\bfseries 92} no.~11,
  (2015) 115148.

\bibitem{tong2018gt}
D.~Tong, ``Gauge theory,'' {\em Lecture notes, DAMTP Cambridge} {\bfseries 10}
  no.~8, (2018) 74. Chapter 4. Lattice Gauge Theory.

\end{thebibliography}\endgroup
	
\end{document}